\documentclass{aa}  

\usepackage{graphicx}           
\usepackage[varg]{txfonts}
\usepackage{subcaption}         
\usepackage{lscape}             
\usepackage{placeins}           
                                
\def\farcs{\hbox{$.\!\!^{\prime\prime}$}}
\def\fsecs{\hbox{$.\!\!^{\mathrm{s}}$}}
\newcommand{\code}[1]{\texttt{#1}}
\newcommand{\CO}{C$^{17}$O}
\newcommand{\ketene}{CH$_2$CO}
\newcommand{\acetaldehyde}{CH$_3$CHO}
\newcommand{\isocyanicacid}{HNCO}
\newcommand{\methylcyanide}{CH$_3$CN}
\newcommand{\cyanoacetylene}{HC$_3$N}
\newcommand{\methanol}{CH$_3$OH}
\newcommand{\formaldehyde}{H$_2$CO}
\newcommand{\Tkin}{$T_{\mathrm{kin}}$}
\newcommand{\nH}{$n_{\mathrm{H_2}}$}
\newcommand{\Tex}{$T_{\mathrm{ex}}$}
\newcommand{\HII}{H\,{\small II}}
\newcommand{\Ha}{H{$\alpha$}}

\usepackage{natbib,twoopt}
\usepackage[breaklinks=true,colorlinks=true,linkcolor=blue]{hyperref} 
\hypersetup{citecolor=blue}
\bibpunct{(}{)}{;}{a}{}{,}             
\makeatletter
  \newcommandtwoopt{\citeads}[3][][]{\href{http://adsabs.harvard.edu/abs/#3}%
    {\def\hyper@linkstart##1##2{}%
     \let\hyper@linkend\@empty\citealp[#1][#2]{#3}}}
  \newcommandtwoopt{\citepads}[3][][]{\href{http://adsabs.harvard.edu/abs/#3}%
    {\def\hyper@linkstart##1##2{}%
     \let\hyper@linkend\@empty\citep[#1][#2]{#3}}}
  \newcommandtwoopt{\citetads}[3][][]{\href{http://adsabs.harvard.edu/abs/#3}%
    {\def\hyper@linkstart##1##2{}%
     \let\hyper@linkend\@empty\citet[#1][#2]{#3}}}
  \newcommandtwoopt{\citeyearads}[3][][]%
    {\href{http://adsabs.harvard.edu/abs/#3}
    {\def\hyper@linkstart##1##2{}%
     \let\hyper@linkend\@empty\citeyear[#1][#2]{#3}}}
\makeatother

\begin{document}

    \title{Spatial distribution of organics in the Horsehead nebula: \\ Signposts of chemistry driven by atomic carbon}
   
   \titlerunning{Spatial distribution of organics in the Horsehead nebula}

   \author{Claudio Hernández-Vera\inst{1,2,3}\fnmsep\thanks{Corresponding author, \email{cahernandezv@uc.cl}}\orcid{0009-0009-2320-7243}
        \and Viviana V. Guzmán\inst{1,2}\orcid{0000-0003-4784-3040}
        \and Jérôme Pety\inst{4,5}\orcid{0000-0003-3061-6546}
        \and Ka Tat Wong\inst{6,4}\orcid{0000-0002-4579-6546}
        \and Javier~R.~Goicoechea\inst{7}\orcid{0000-0001-7046-4319}
        \and Franck Le Petit\inst{8}\orcid{0000-0001-8738-6724}
        \and Maryvonne Gerin\inst{5}\orcid{0000-0002-2418-7952}
        \and Aquiles den Braber\inst{1}\orcid{0009-0002-7819-6982}
        \and John M. Carpenter\inst{9}\orcid{0000-0003-2251-0602}
        \and Vincent Maillard \inst{7}\orcid{0009-0003-1327-3737}
        \and Emeric~Bron\inst{8}\orcid{0000-0003-1532-7818}
        \and Pierre~Gratier\inst{10}\orcid{0000-0002-6636-4304}
        \and Evelyne Roueff\inst{8}\orcid{0000-0002-4949-8562}
        }

   \institute{Instituto de Astrof\'isica, Pontificia Universidad Cat\'olica de Chile, Av. Vicu\~na Mackenna 4860, 7820436 Macul, Santiago, Chile
    \and Millennium Nucleus on Young Exoplanets and their Moons (YEMS)
    \and European Southern Observatory, Alonso de Córdova 3107, Casilla 19, Vitacura, Santiago, Chile
    \and IRAM, 300 rue de la Piscine, 38400 Saint Martin d'H\`eres, France
    \and LUX, Observatoire de Paris, PSL Research University, CNRS, Sorbonne Universités, 75014 Paris, France
    \and Theoretical Astrophysics, Department of Physics and Astronomy, Uppsala University, Box 516, 75120 Uppsala, Sweden
    \and Instituto de Física Fundamental (CSIC), Calle Serrano 121, 28006, Madrid, Spain
    \and LUX, Observatoire de Paris, PSL Research University, CNRS, Sorbonne Universités, 92190 Meudon, France
    \and Joint ALMA Observatory, Avenida Alonso de Córdova 3107, Vitacura, Santiago, Chile
    \and Laboratoire d’Astrophysique de Bordeaux, Univ. Bordeaux, CNRS, B18N, Allée Geoffroy Saint-Hilaire, 33615 Pessac, France
    }

   \date{Received 25 September 2025 /
Accepted 27 January 2026}

\abstract{Complex organic molecules (COMs) are considered essential precursors to prebiotic species in the interstellar and circumstellar medium. Despite their astrobiological relevance, many aspects of  the formation of COMs  remain unclear, particularly the role of ultraviolet (UV) radiation. While COMs were once expected to be efficiently destroyed under UV-irradiated conditions, detections in photodissociation regions (PDRs) have challenged this view. However, the mechanisms by which UV radiation contributes to their formation are still uncertain. Here we present moderately resolved maps of simple and complex organic molecules at the UV-illuminated edge of the Horsehead nebula, obtained by combining Atacama Large Millimeter/submillimeter Array (ALMA) and IRAM 30m single-dish observations at $\sim$$15^{\prime\prime}$ resolution. For the first time in this PDR environment, we analyzed the spatial distribution of species such as C$^{17}$O, CH$_2$CO, CH$_3$CHO, HNCO, CH$_3$CN, and HC$_3$N. By incorporating previous C$^{17}$O and C$^{18}$O single-dish data as well as Plateau de Bure Interferometer (PdBI) maps of H$_2$CO and CH$_3$OH, we derived profiles of gas density, temperature, thermal pressure, and column densities of the organic species as a function of distance from the UV source. Our results show that most organic species—particularly H$_2$CO, CH$_2$CO, CH$_3$CHO, HNCO, and CH$_3$CN—exhibit enhanced column densities at the UV-illuminated edge compared to cloud interiors, possibly indicating efficient dust-grain surface chemistry driven by the diffusion of atomic C and radicals produced via photodissociation of CO and CH$_3$OH, as supported by recent laboratory experiments. The exceptions, HC$_3$N and CH$_3$OH, can be attributed to inefficient formation on dust grains and ineffective nonthermal desorption into the gas phase, respectively. Additionally, contributions from gas-phase hydrocarbon photochemistry, possibly seeded by grain-surface products, cannot be ruled out. Further chemical modeling is needed to confirm the efficiency of these pathways for the studied species, which could have important implications for other cold UV-irradiated environments such as protoplanetary disks.}


\maketitle 

\begin{figure*}[t!]
\sidecaption
\includegraphics[width=11.0cm]{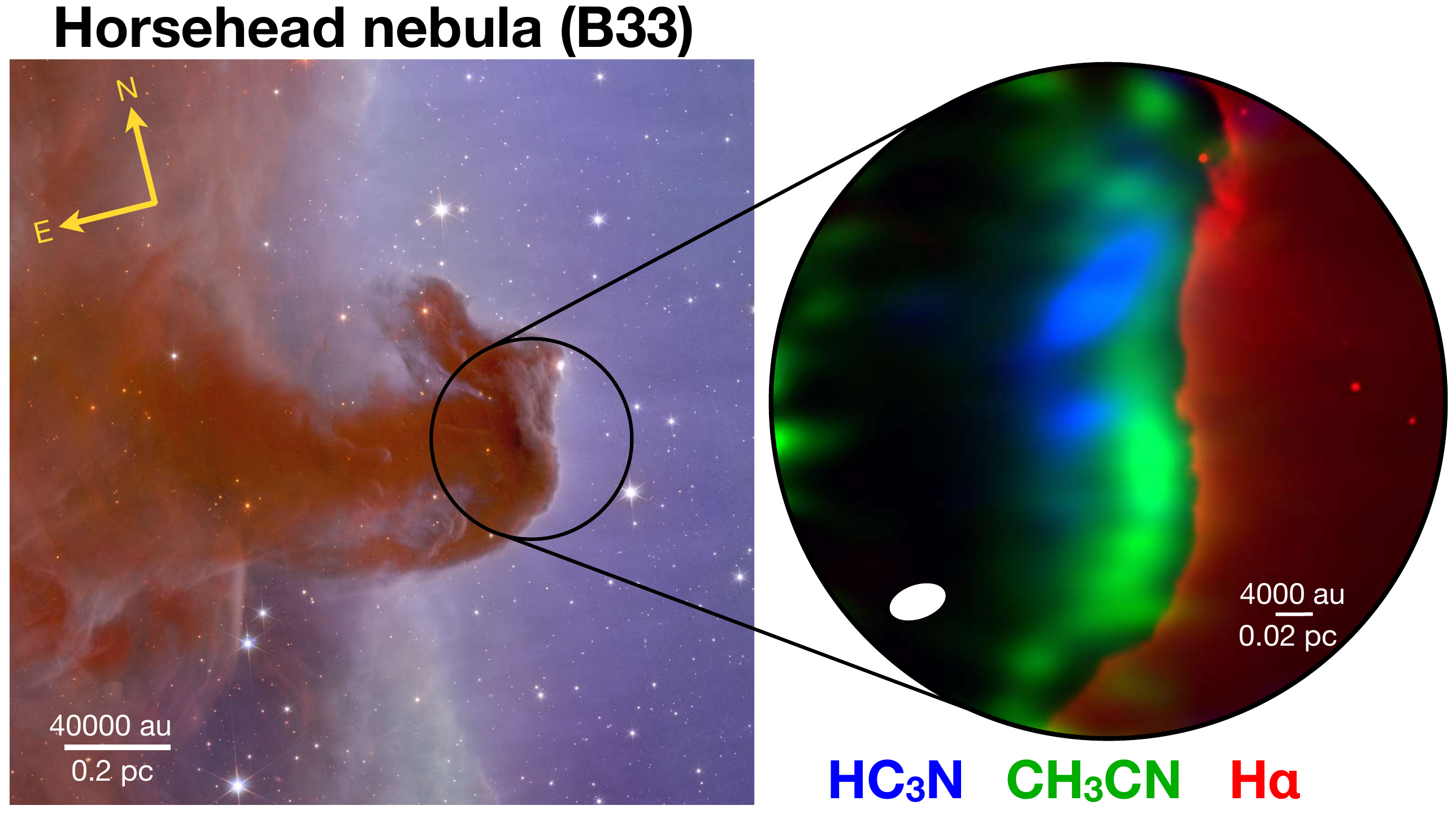}
\caption{Composite image of the field of view mapped in the Horsehead nebula (Barnard 33). \textit{Left}: Horsehead nebula and  adjacent \HII{} region IC 434 imaged by Euclid Early Release Observations (ESA/Euclid/Euclid Consortium/NASA). \textit{Right}: Zoomed-in image of the edge of the Horsehead nebula imaged with data combined from the ALMA 7m and the IRAM 30m telescopes. The emission of different N-bearing molecules are represented:  \methylcyanide{} ($J_{K} = 6_{0}-5_{0}$, in green) and \cyanoacetylene{} ($J = 11-10$, in blue). The emission from hot ionized gas is also shown, traced by the \Ha{} line (in red) observed with the $0.9$~m KPNO telescope \citep{Pound2003}. The average beam size of the ALMA 7m+30m observations and a physical scale reference are shown in the bottom left and bottom right corner, respectively.}
\label{fig:hh-Ncoms-rgb}
\end{figure*}

\section{Introduction}\label{Sec:Intro}

Molecular line observations provide a powerful tool for probing the structure and physical conditions of the molecular gas across a wide range of environments, from high-redshift galaxies \citep{Carilli2013} to protoplanetary disks \citep{Oberg2023}. To date, about 330 molecules have been detected in the interstellar medium (ISM) and in circumstellar shells.\footnote{https://cdms.astro.uni-koeln.de/classic/molecules} While most detections involve simple diatomic or triatomic species, a significant subset consists of organic molecules with six or more atoms, loosely defined as complex organic molecules \citep[COMs;][]{Herbst2009}. 

The study of COMs and their precursors is critical not only because they trace the physical and chemical evolution across different stages of star and planet formation, but also because they are potential precursors of prebiotic species, the building blocks of terrestrial life \citep{Ceccarelli2023}. Fundamental compounds such as amino acids and nucleotide bases are believed to have an interstellar origin \citep{Garrod2013,Altwegg2016,Jimenez-Serra2020,Rivilla2019,Rivilla2023}, though their formation pathways remain uncertain. Therefore, dedicated studies of the formation mechanisms of simpler COMs are essential to clarify their role in producing the ingredients necessary for life as we know it.

The origin of COMs remains an active area of investigation in astrochemistry. Traditionally, COMs have been abundantly detected in hot cores around high-mass protostars \citep[e.g.,][]{Bisschop2007,Feng2015,Bonfand2017,vanderwalt2021} and in their low-mass counterparts, hot corinos \citep[e.g.,][]{Cazaux2003,Oberg2014,Belloche2020,Yang2021}. Their formation was typically attributed to small organics produced via grain-surface reactions from inherited ices, which are then thermally desorbed and processed further through gas-phase chemistry \citep{Herbst2009}. However, the detection of abundant COMs in pre-stellar cores has challenged this view \citep[e.g.,][]{Bacmann2012,Cernicharo2012,Jimenez-Serra2016,Scibelli2024}, highlighting the possible importance of cold gas-phase reactions \citep{Shannon2013}, nondiffusive chemistry on grains \citep{Jin2020,Garrod2022}, and nonthermal desorption processes such as chemical desorption \citep{Garrod2007}.

Thus, there is general consensus that the observed abundances of COMs likely arise from an intricate interaction between gas-phase and grain-surface chemistry \citep{Jorgensen2020,Ceccarelli2023}, though many aspects remain poorly understood. In particular, the role of ultraviolet (UV) radiation is crucial as it can further process interstellar ices and promote their release into the gas phase through nonthermal photodesorption \citep{Oberg2016}. Moreover, laboratory experiments demonstrate that UV irradiation of ices can even lead to the formation of prebiotic species \citep{deMarcellus2015,Oba2019}. These processes are especially relevant in environments such as protoplanetary disks, where compelling evidence for in situ ice processing has recently been proposed \citep[e.g.,][]{Walsh2016,Calahan2023,Ligterink2024,Yamato2024,Evans2025}. However, the generally low abundance of COMs and the complex physicochemical structure of these environments \citep[see][for a recent review]{Oberg2023} make their study especially challenging.

A promising way to investigate the impact of UV radiation on COM chemistry is to study the edges of molecular clouds illuminated by massive stars, known as the photodissociation regions \citep[PDRs;][]{Hollenbach1999}. In these environments, far-UV (FUV; $6 < h \nu < 13.6$~eV) photons dominate the chemistry and were initially thought to efficiently photodissociate large molecules. However, more recent studies have challenged this view \citep[see][for a recent review]{Wolfire2022}. Particularly, COMs have been observed to be abundant in PDRs \citep[e.g.,][]{Guzman2014,Cuadrado2017}, even when the radiation field exceeds the interstellar average by several orders of magnitude \citep[$G_0 \sim 10^2-10^4$, where $G_0 = 1.7$ is the average interstellar radiation field;][]{Draine1978}. Pure gas-phase models cannot account for the observed high COM abundances in these regions \citep[e.g.,][]{Gratier2013,Guzman2013,Cuadrado2017}, pointing to a key role for FUV processing of ices. However, the details of this chemistry remain poorly constrained observationally: only a handful of targeted studies exist \citep[e.g.,][]{Guzman2014,Cuadrado2017}, all relying on single-dish telescopes, and thus lacking spatial resolution to determine how COMs are distributed relative to the UV source.

Illuminated by the O9.5V binary system $\sigma$~Orionis \citep{Warren1977} $\sim$$3.5$~pc away from the molecular cloud \citep{Abergel2003}, the PDR located at the edge of the Horsehead nebula is particularly well-suited for a spatially resolved study. Thanks to its proximity \citep[$\sim$$400$~pc,][]{Anthony-Twarog1982}, nearly edge-on geometry, and moderate radiation field \citep[$G_0\sim100$,][]{Habart2005}, the Horsehead PDR is an archetype for the average UV radiation fields in molecular clouds of the Milky Way and normal star-forming galaxies. As a consequence, its gas and dust content have been thoroughly characterized at different wavelengths \citep[][and references therein]{Schirmer2020,Hernandez-Vera2023,Abergel2024}. Specifically, the Horsehead  Wideband High-resolution IRAM 30m Surveys at two Positions with EMIR Receivers (WHISPER, PI: J. Pety) single-dish line survey provided a deep chemical inventory at two key locations:\footnote{The PDR is located at $\alpha_{2000} = 05^\mathrm{h}40^\mathrm{m}53\fsecs936$, $\delta_{2000} = -02^\circ28^\prime00^{\prime\prime}$ and the dense core is located at $\alpha_{2000} = 05^\mathrm{h}40^\mathrm{m}55\fsecs61$, $\delta_{2000} = -02^\circ27^\prime38^{\prime\prime}$.} the HCO emission peak \citep[PDR position,][]{Gerin2009} and the DCO$^{+}$ emission peak \citep[dense core position,][]{Pety2007}. Strikingly, this survey yielded the first PDR detections of complex species such as HCOOH, CH$_2$CO, CH$_3$CHO, and CH$_3$CCH \citep{Guzman2014}. Since then, several smaller molecules have been mapped at higher angular resolution with the Plateau de Bure Interferometer \citep[PdBI; e.g.,][]{Guzman2013,Guzman2015} and the Atacama Large Millimeter/submillimeter Array \citep[ALMA; e.g.,][]{Hernandez-Vera2023}, though similar mapping of COMs remains scarce.

For the present study, we investigated the spatial distribution of O-bearing and N-bearing organics at the UV-illuminated edge of the Horsehead nebula, combining ALMA and IRAM 30m data to achieve $\sim$$15^{\prime\prime}$ resolution (see Fig.~\ref{fig:hh-Ncoms-rgb} for an overview). We examined how the physical conditions of gas vary across the cloud and explored the chemical formation pathways of COMs under moderate FUV irradiation. Section~\ref{Sec:Obs} describes the observations and data combination. Section~\ref{Sec:Results} presents the spatial distributions, physical conditions, and column densities of the analyzed species. In Sect.~\ref{Sec:discussion} we elaborate on the possible formation pathways of the studied molecules and discuss implications for FUV-driven organic chemistry in analogous environments. The main results and our conclusions are summarized in Sect.~\ref{Sec:summary-conclusions}.

\section{Observations and data reduction}\label{Sec:Obs}

In this section, we describe the interferometric ALMA deep observations of
simple organics and COMs taken at the edge of the Horsehead nebula that
were combined with matched IRAM 30m single-dish observations. For the analysis,
they were complemented with Plateau de Bure Interferometer (PdBI) legacy
data, previously published by \cite{Guzman2013}, and the publicly available WHISPER data.

\subsection{ALMA 7m observations}\label{Subsec:ACA-obs}

We used the ALMA Atacama Compact Array (hereafter ALMA 7m) to perform deep observations of the Horsehead nebula at $\sim 15^{\prime\prime}$ between August 1, 2017, and October 1, 2017, during Cycle 4 (2016.2.00027.S, PI: V.V. Guzmán). There were 22 successful execution blocks (EBs), each lasting for about $1.5$~h, with an on-source time of $0.8$~h. Single pointing observations centered at $\alpha_{2000} = 05^\mathrm{h}40^\mathrm{m}54\fsecs555$, $\delta_{2000} = -02^\circ27^\prime48\farcs6$ were carried out in Band 3 with projected baseline lengths between 5 and 49 m for an average of ten antennas per EB. The primary beamwidth at $100$~GHz is $107.9^{\prime\prime}$ (see Fig.~\ref{fig:hh-Ncoms-rgb}, for a reference in physical units, assuming a distance of $400$~pc). Quasar J0510$+$1800 or J0522$-$3627 was observed as the bandpass calibrator; J0423$-$0120, J0522$-$3627, or Uranus as the absolute flux calibrator; and J0541$-$0211 as the phase calibrator. 

A single spectral setup was defined, covering the $98-113$~GHz range and divided into nine spectral windows in the frequency division mode (FDM) of the correlator. As listed in Table~\ref{table:aca}, these windows target the continuum at $3$~mm and the following rotational transitions: carbon monoxide (\CO{}) $J = 1-0$, ketene (\ketene{}) $J_{K_a,K_c}=5_{15}-4_{14}$, acetaldehyde (\acetaldehyde{}) $J_{K_a,K_c}=5_{14}-4_{13}$ (only E symmetry) and $6_{16}-5_{15}$ (both E and A symmetries), isocyanic acid (\isocyanicacid{}) $J_{K_a,K_c}=5_{05}-4_{04}$, cyanoacetylene (\cyanoacetylene{}) $J = 11-10$, and methyl cyanide (\methylcyanide{}) $J_{K} = 6_{0}-5_{0}$ (A), $6_{1}-5_{1}$ (E), and $6_{2}-5_{2}$ (E). Other molecules such as methyl isocyanide (CH$_3$NC) and propynylidyne ($l$-C$_3$H) were also targeted. The former was not detected, whereas the latter was, but its analysis lies beyond the scope of this work.

Due to the small aggregate bandwidth, the bandwidth-switching technique was adopted in order to improve the signal-to-noise ratios of the flux and gain calibration. Another spectral setup was defined for the calibrator scans. The calibration setup includes four wide-band spectral windows in the time division mode (TDM), centering at roughly $98.32$, $100.28$, $110.32$, and $112.32$~GHz with the channel spacing of $15.625$~MHz and the bandwidth of $2.0$~GHz. In each EB, the bandpass calibrator\footnote{This is now referred to as the differential gain calibrator and is observed with the DIFFGAIN scan intent in recent cycles.} was observed in both FDM (narrow-band) and TDM (wide-band) setups to determine the phase differences between the FDM and TDM windows \citep[see, e.g., ALMA Technical Handbook for details][]{almathb2025}, while the flux and phase calibrators were observed with the TDM setup only.

The data were initially calibrated manually at the Joint ALMA Observatory (JAO) using the Common Astronomy Software Application (\code{CASA})\footnote{https://casa.nrao.edu/} version 5.1.1 \citep{McMullin2007,CASA2022} with the calibration scripts created by the ALMA Script Generator \citep{Petry2014}. We have modified the scripts and recalibrated the data with \code{CASA} version 5.4.0 in order to optimize the spectral window mapping schemes for the system temperature ($T_{\rm sys}$) and gain (amplitude and phase) calibrations. The $T_{\rm sys}$ of the calibrators and science target data were derived from the measurement scans of the same source and spectral window. The gain solutions for the narrow-band windows of the target were derived from the wide-band windows of the phase calibrator in the same baseband (see Table~\ref{table:aca}). In addition, antenna CM06 was manually flagged due to unreliable $T_{\rm sys}$ values. After flagging, the on-source visibilities amount to about 14 hours of integration times with ten antennas.

\begin{table*}
\caption{Observational parameters for the ALMA 7m+30m maps shown in Figs.~\ref{fig:hh-coms-grid} and \ref{fig:hh-c17o-grid}.}              
\label{table:obs-params}
\centering
\begin{tabular}{l c c c c c c}          
\hline\hline
Species & Transition & Chan. Width & Beam & Beam PA & Chan. rms & Mom. Zero rms \\ &  & (km~s$^{-1}$) & ($^{\prime\prime}$) & $^\circ$ & (mK) & (mK km~s$^{-1}$)\\
\hline                                   
C$^{17}$O & $J=1-0$ & $0.20$ & $15.69$ $\times$ $8.79$ & $-73.43$ & $55.52$ & $27.20$\\
CH$_2$CO & $J_{K_a,K_c}=5_{15}-4_{14}$ & $0.20$ & $17.08$ $\times$ $9.70$ & $-77.77$ & $34.00$ & $16.65$ \\
CH$_3$CHO & $J_{K_a,K_c}=5_{14}-4_{13}$ (E) & $0.20$ & $17.28$ $\times$ $9.85$ & $-76.92$ & $29.26$ & $14.33$ \\
 & $J_{K_a,K_c}=6_{16}-5_{15}$ (E) & $0.20$ & $15.68$ $\times$ $8.80$ & $-73.36$ & $47.74$ & $21.35$ \\
 & $J_{K_a,K_c}=6_{16}-5_{15}$ (A) & $0.20$ & $15.68$ $\times$ $8.80$ & $-73.36$ & $47.74$ & $21.35$ \\
HNCO & $J_{K_a,K_c}=5_{05}-4_{04}$ & $0.20$ & $15.62$ $\times$ $9.01$ & $-77.83$ & $53.71$ & $26.31$ \\
HC$_3$N & $J=11-10$ & $0.20$ & $17.08$ $\times$ $9.70$ & $-77.78$ & $35.82$ & $17.55$ \\
CH$_3$CN & $J_K = 6_{0}-5_{0}$ (A) & $0.20$ & $15.52$ $\times$ $8.91$ & $-78.00$ & $63.00$ & $35.64$ \\
 & $J_K = 6_{1}-5_{1}$ (E) & $0.20$ & $15.52$ $\times$ $8.91$ & $-78.00$ & $63.00$ & $35.64$ \\
 & $J_K = 6_{2}-5_{2}$ (E) & $0.20$ & $15.52$ $\times$ $8.91$ & $-78.00$ & $63.00$ & $35.64$ \\
\hline
\end{tabular}
\tablefoot{The projection center of the maps is $\alpha_{2000} = 05^\mathrm{h}40^\mathrm{m}54\fsecs555$, $\delta_{2000} = -02^\circ27^\prime48\farcs6$
}
\end{table*}

\subsection{IRAM 30m observations}\label{Subsec:IRAM30m-obs}

As part of project 108-17, we used the EMIR receivers of the IRAM 30m telescope to map during 47 hours a region of $180^{\prime\prime} \times 180^{\prime\prime}$ centered on ALMA 7m pointing and rotated by $14^\circ$ counterclockwise compared to
the Equatorial $2000$ coordinate frame. We used the narrow mode of the Fourier Transform Spectrometer that delivers four times $1.8$~GHz of continuous bandwidths (two of them in each sideband) at a channel spacing of $50$~kHz. We interleaved two different receiver tunings to cover continuously from the lower sideband from $93.65$ to $100.55$~GHz, and the upper sideband from $109.65$ to $116.65$~GHz. We used the position-switch, on-the-fly observing mode, covering each field with back-and-forth scans along either the right ascension or declination axes. We slewed at a speed
of $8.5^{\prime\prime}$~s$^{-1}$ and we dumped data to disk every $0.5$~s, yielding at least 5 integrations per beam in the scanning direction. The scan legs were separated by $8.5^{\prime\prime}$, yielding Nyquist sampling transverse to the scan
direction at $2.6$~mm. We calibrated the system every $15$ minutes.

The calibration of the data was done inside the \code{GILDAS/MRTCAL} software and the remaining of the processing happened in the \code{GILDAS/CLASS} software. Around the frequency of each line of interest, we extracted a frequency range of $100$~MHz. We baselined the spectra by fitting a polynomial of order 1 after excluding the [$8.5$, $12.5$~km~s$^{-1}$] velocity range. We converted the spectra intensity in main beam temperature using the standard Ruze formula for the IRAM 30m telescope. Finally we resampled the velocity axis to match the ALMA 7m axis, and we gridded the data in a cube of $4^{\prime\prime}\times 4^{\prime\prime}$ pixels.

\subsection{Interferometric and single-dish data combination}\label{Subsec:data-comb}

We combined the IRAM 30m and ALMA 7m to produce a clean image containing all spatial information from $0$ to $50$~m, i.e., equivalent to a $50$~m diameter total-power telescope. The IRAM 30m map is deconvolved from the $30$~m beam in the Fourier plane before multiplication by the ALMA 7m primary beam in the image plane. After a last Fourier transform, pseudo-visibilities were sampled between $0$ and $7$~m. These visibilities were then merged with the ALMA 7m interferometric observations \citep[see][for details]{Rodriguez08}. Standard imaging, CLEAN deconvolution, and primary beam correction steps were then applied. The resulting data cube was then scaled from Jy~beam$^{-1}$ to $T_{\mathrm{mb}}$ temperature scale using the synthesized beam size. All these processes happened in the \code{GILDAS/MAPPING} software.

The observational parameters of the final combined (hereafter ALMA 7m+30m) images are summarized in Table~\ref{table:obs-params}. The channel rms is calculated using the central 50\% of the pixels in the first and last five (line-free) channels of the data cubes. To compare the spatial distribution of the different molecular tracers, we collapse the line emission into velocity-integrated intensity, or \textit{zeroth-moment}, maps using the \code{CASA} task \code{immoments}. The zeroth-moment rms is estimated as $\sigma_{\mathrm{rms}} \approx \sigma_{\mathrm{chan}} \sqrt{N_{\mathrm{chan}}} \Delta v$, where $\sigma_{\mathrm{chan}}$ is the channel rms, $N_{\mathrm{chan}}$ is the number of channels used to compute the zeroth-moment, and $\Delta v$ is the channel spacing in velocity.

\subsection{Legacy data}\label{Subsec:auxlliary-data}

The analysis of the ALMA 7m+30m data was complemented with PdBI observations of formaldehyde (\formaldehyde{}) $J_{K_a,K_c}=2_{02}-1_{01}$ and methanol (\methanol{}) $J_{K} = 3_{-1}-2_{-1}$ (E symmetry) rotational transitions imaged at $\theta_{\mathrm{beam}}\sim6^{\prime\prime}$. All the observational details of these legacy data can be found in \cite{Guzman2013}. In order to make the two data sets comparable, the PdBI data was rotated by 14$^\circ$, and the angular resolution was downgraded to $\theta_{\mathrm{beam}}\sim15^{\prime\prime}$ to match the ALMA 7m+30m data using the \code{CASA} tools \code{rotate} and \code{convolve2d}, respectively. Afterward, the spatial axes were regridded using the \code{CASA} task \code{imregrid} to match the number of pixels per beam. Throughout this procedure, the CH$_3$CN image was used as a template. Thus, for all subsequent analysis, the angular resolution of the H$_2$CO and CH$_3$OH images is exactly the same as for CH$_3$CN (see Table~\ref{table:obs-params}).

Additionally, we extracted from the WHISPER line survey\footnote{https://iram.fr/\textasciitilde pety/horsehead/Horsehead\_Nebula/Data.html} the spectra of the $J=1-0$ ($\theta_{\mathrm{beam}} \sim 22^{\prime\prime}$) and $2-1$ ($\theta_{\mathrm{beam}} \sim 11^{\prime\prime}$) rotational transitions for the C$^{17}$O and C$^{18}$O isotopologues at the PDR and dense core positions, to estimate the $\mathrm{C}^{17}\mathrm{O}/\mathrm{C}^{18}\mathrm{O}$ ratio in the Horsehead nebula and find the excitation temperature of C$^{17}$O (see Sect.~\ref{Subsubsec:conversion-factor} and Appendix~\ref{subApp:WHISPER}). The observational details of WHISPER can be found in any of the survey papers \citep[e.g.,][]{Guzman2012,Pety2012}.

\begin{figure*}[t!]
\centering
\includegraphics[width=0.9\linewidth]{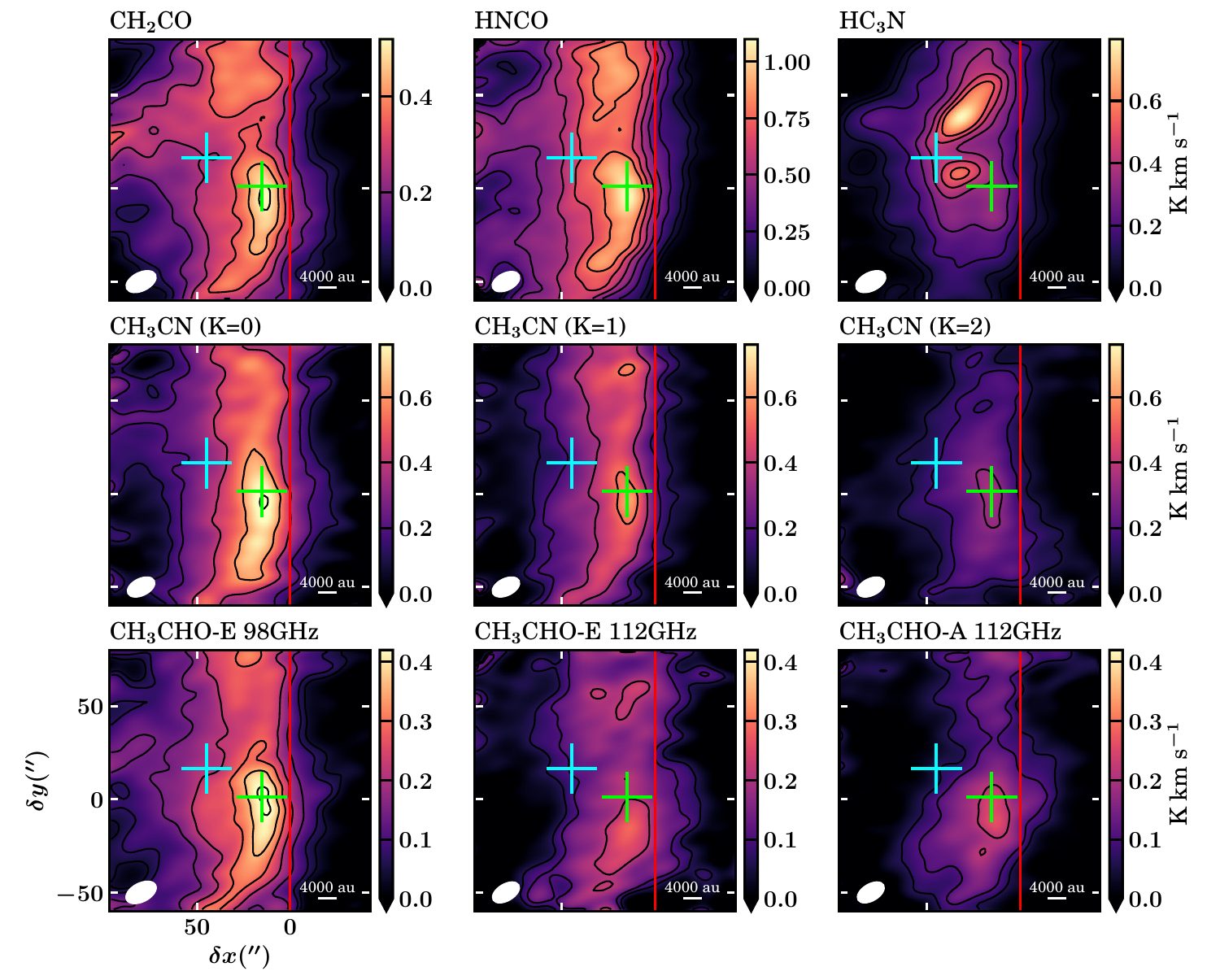}
\caption{Zeroth-moment maps gallery for different molecular species moderately resolved at the Horsehead edge. The maps have been rotated $14^\circ$ counterclockwise to bring the illuminating star direction in the horizontal direction. The contours are $[3, 5, 10, 15, 20, 25, 30]\times\sigma$, where $\sigma$ is the zeroth-moment rms listed in Table~\ref{table:obs-params}. The red vertical line represents the horizontal zero, delineating the PDR edge \citep{Pety2005}, whereas the cyan and green crosses show the dense core \citep{Pety2007} and PDR \citep{Gerin2009} positions, respectively. The beam size and a scale bar indicating $4000$~au are shown in the bottom left and bottom right corner, respectively, of each panel. Transitions corresponding to the same molecular species are shown using the same color-scale range.}
\label{fig:hh-coms-grid}
\end{figure*}

\begin{figure*}[t!]
\sidecaption
\includegraphics[width=12cm]{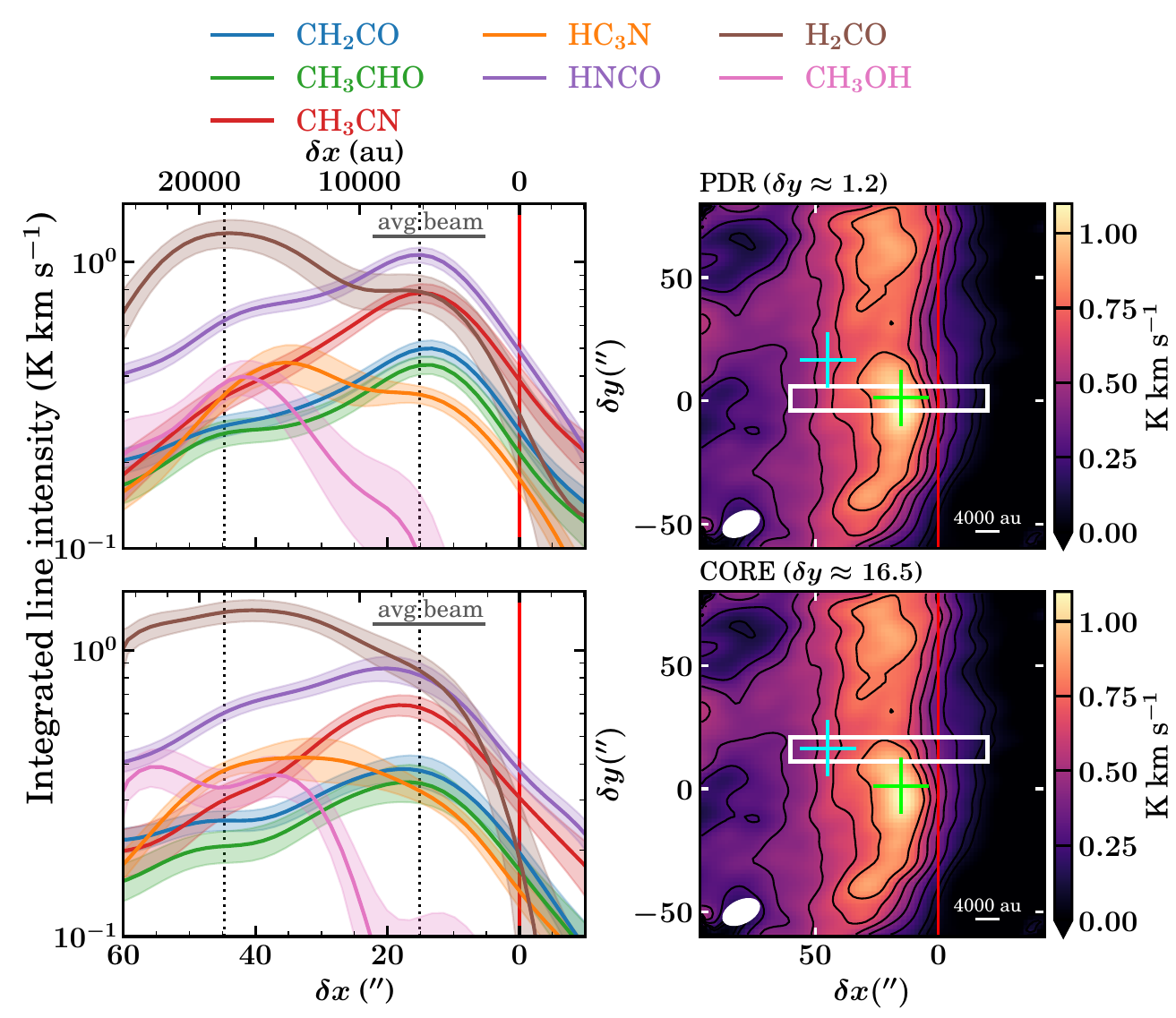}
\caption{\textit{Left column}: Integrated line intensity profiles along the direction of the exciting star extracted from the maps shown in Fig.~\ref{fig:hh-coms-grid} and the convolved maps obtained from \cite{Guzman2013}. For molecules with more than one transition, the brightest is displayed. The profiles were extracted at the $\delta y$ position of the PDR (top panel) and dense core (bottom panel), and averaged over $10^{\prime\prime}$ in the $\delta y$ direction. The dotted vertical lines represent the $\delta x$ position of the PDR and dense core, and the vertical red line represents the PDR edge. The average beam size is represented by the horizontal gray bar. The colored area of each profile displays the $\pm \sigma$ significance levels, taking into account the standard deviation of the average and the zeroth-moment rms. \textit{Right column}: Regions used to extract the profiles, indicated by white rectangles overlaid on the HNCO zeroth-moment map. The two crosses, the red vertical line, and the contours are the same as in Fig.~\ref{fig:hh-coms-grid}.}
\label{fig:hh-coms-profiles}
\end{figure*}

\section{Results}\label{Sec:Results}

This section presents the main findings from our combined ALMA 7m+30m observations. We describe the spatial distribution of simple and complex organics in the Horsehead nebula, based on integrated intensity maps and their corresponding horizontal profiles at key vertical positions. Next, we detail the radiative transfer modeling used to constrain the physical conditions and derive column density profiles for each organic species. Finally, we compare our results with previous studies in the same source and discuss the differences and similarities among the species analyzed.

\subsection{Spatial distribution of molecules}\label{Subsec:spatial-distrib}

Figure~\ref{fig:hh-coms-grid} shows the gallery of zeroth-moment maps for all the molecular species listed in Table~\ref{table:obs-params}, except for C$^{17}$O that is shown in Fig.~\ref{fig:hh-c17o-grid}. Following the usual conventions of previous work, the maps were rotated $14^\circ$ counterclockwise to bring the exciting star direction in the horizontal $\delta x$ direction. By doing this, a new rotated coordinate system is defined
where the horizontal zero ($\delta x = 0$) corresponds to the empirical PDR edge (red line, Fig.~\ref{fig:hh-coms-grid}) usually defined at the sharp boundary of the molecular cloud traced by the H$_2$ ro-vibrational line emission \citep{Pety2005}. Using the rotated coordinates, the PDR (green cross, Fig.~\ref{fig:hh-coms-grid}) and dense core (cyan cross, Fig.~\ref{fig:hh-coms-grid}) are located at ($\delta x \approx 15\farcs1,\:\delta y \approx 1\farcs2$) and ($\delta x \approx 44\farcs7,\:\delta y \approx 16\farcs5$), respectively. 

The visual inspection of Fig.~\ref{fig:hh-coms-grid} reveals a distinct trend: most molecules, except for \cyanoacetylene, show a filament structure whose emission peak is at the PDR position. However, it is also clear that O-bearing and N-bearing molecules apparently have different spatial distributions, which probably implies disparities in their chemistry (see Sect.~\ref{Sec:discussion}, for a detailed discussion). In addition to exhibiting a filament at the PDR, O-bearing species, such as \ketene{} and \acetaldehyde{}, have an extended component towards the position of the dense core, following the shape of the ``Horsehead pillar" typically shaped by large-scale structure maps of CO isotopologues \citep[][see also Fig.~\ref{fig:hh-c17o-grid}]{Roueff2024,Segal2024}. Instead, N-bearing species, such as \methylcyanide{} and \cyanoacetylene{}, are notably different. \methylcyanide{} lacks significant extended emission at the dense core and the PDR filament is thinner than the O-bearing ones. In contrast, \cyanoacetylene{} reveals a more clumpy morphology concentrated close to the dense core. In between, we have HNCO, exhibiting characteristics of both types of species: a sharply defined filament across the PDR along with some extended emission near the dense core, consistent with its dual N and O composition.

Interestingly, both the filament centered at the PDR and the clumps close to the dense core have been previously seen by higher angular resolution data of other molecules. \cite{Guzman2013} showed that \formaldehyde{} emission presents a similar morphology, and suggested it is a consequence of dust-grain versus gas-phase chemistry at the PDR and dense core, respectively. Consistently, \cite{Hernandez-Vera2023} concluded that HCO$^+$ exhibits a similar type of distribution since it traces the dense UV-shielded gas but also the more diffuse edge of the molecular cloud exposed to stellar UV radiation. Thus, the different distributions of \methylcyanide{} and \cyanoacetylene{} seen in the maps could be direct evidence of different formation pathways, as previously suggested by \cite{Gratier2013}, or variations on the gas-phase C/O ratio, as mentioned by \cite{LeGal2019}. Likewise, similarities between \ketene{} and \acetaldehyde{} might be associated with common formation routes, as proposed by \cite{Guzman2014}. More details on the possible formation pathways are discussed in Sect.~\ref{Sec:discussion}.

Similar to previous works \citep[e.g.,][]{Guzman2015,Hernandez-Vera2023}, we extracted horizontal cuts from the maps to quantify the distribution of the line emission as a function of $\delta x$, particularly at the $\delta y$ position of the PDR and dense core. In this case, we averaged the integrated line intensity over $10^{\prime\prime}$ in the $\delta y$ direction since it roughly matches the beam of the observations in that direction (see right panels, Fig.~\ref{fig:hh-coms-profiles}). Hereafter, we decided to include the convolved maps of \formaldehyde{} and \methanol{} in the analysis to compare the emission of these smaller species with the larger COMs presented here. Moreover, they can be essential precursors of more complex species \citep{Oberg2009,Lopez-Sepulcre2024}.

Figure~\ref{fig:hh-coms-profiles} shows the integrated intensity profiles extracted at $\delta y \approx 1\farcs2$ (PDR) and $\delta y \approx 16\farcs5$ (dense core). The uncertainties of each profile incorporate the standard deviation of the average along the $\delta y$ direction and the zeroth-moment rms from Table~\ref{table:obs-params}, added in quadrature. We also accounted for an absolute flux calibration uncertainty of approximately 5\%. For \formaldehyde{} and \methanol{}, the zeroth-moment rms is estimated from the observational parameters reported by \cite{Guzman2013}. It is now more evident that most molecules shown in Fig.~\ref{fig:hh-coms-grid} have their emission peak at the PDR filament and that the N-bearing molecules \methylcyanide{} and \cyanoacetylene{} have different spatial distributions. Hence, to further explore the above, we used the integrated intensity profiles to determine the gas physical conditions and the column densities profiles for each molecule.

\subsection{Radiative transfer modeling}\label{Subsec:radiative-transfer}

We have performed radiative transfer calculations to model the line intensities shown in Fig.~\ref{fig:hh-coms-profiles}. Typically, the usual way of doing this is through a rotational diagram analysis \citep{Goldsmith1999}, which assumes optically thin emission and level populations following a Boltzmann distribution characterized by a single excitation temperature (\Tex{}) and the total column density of the molecule ($N$). Under local thermodynamic equilibrium (LTE), \Tex{} provides a good approximation of the gas kinetic temperature (\Tkin{}). However, previous studies have shown that complex molecules in the Horsehead edge are, in general, subthermally excited \citep[$T_{\mathrm{ex}}<T_{\mathrm{kin}}$; see][and references therein]{Guzman2014}. Therefore, to obtain more accurate results, it is always desirable to take a non-LTE approach, in which both collisional and radiative de-excitation must be considered.

We used the non-LTE radiative transfer code RADEX \citep{vanderTak2007}. More specifically, we ran our analyses using the Python-wrapped version \code{SpectralRadex\footnote{https://github.com/uclchem/SpectralRadex}} \citep{Holdship2021}. Taking this approach, the modeled emission of a specific molecule (in units of K km s$^{-1}$) will depend on three free parameters: the gas density of the collisional partners, typically H$_2$ (\nH{}),\footnote{We neglect the possible influence of H, He, and e$^-$ as collisional partners.} the gas temperature (\Tkin{}), and the total column density of the molecule ($N$). Other parameters, such as the background temperature ($T_{\mathrm{bg}}$) and the linewidth ($\Delta v$), can also be varied. For simplicity, and because they are not expected to vary significantly across the analyzed region, we kept them fixed at $T_{\mathrm{bg}} = 2.73$~K (CMB temperature) and $\Delta v = 0.6$~km~s$^{-1}$, as supported by previous studies \citep{Gratier2013,Guzman2014}.

We fitted the modeled emission to the observed integrated intensities by constructing a likelihood function, $\mathcal{L}(\mathrm{data}, \mathrm{free\:parameters})$, and performing a $\chi^2$ minimization. To retrieve the posterior probability distributions of the free model parameters, we employed the affine-invariant Markov chain Monte Carlo (MCMC) algorithm implemented in the \code{emcee} package \citep{Foreman-Mackey2013}. The number of walkers was set to four times the number of free model parameters, with 2000 burn-in steps and an additional 3000 steps used to sample the posterior distributions. The observational uncertainties used in the fit correspond to those shown in the profiles in Fig.~\ref{fig:hh-coms-profiles}. The best-fit values and their associated uncertainties were derived from the 16th, 50th, and 84th percentiles of the posterior distributions.

The number of free model parameters in the fit depended on the number of observed transitions available for each molecule. For \methylcyanide{}, which has multiple detected rotational lines, we allowed \Tkin{} and $N$ to vary freely. For the other molecules, which had only one or two detected lines, we fixed some of the physical parameters (typically \nH{} and/or \Tkin{}) based on the results from \methylcyanide{} or literature values, and then constrained the column density $N$ through the fitting. To avoid oversampling given the $\sim$$15^{\prime\prime}$ beam size, the full analysis was carried out at intervals of $\sim$$7.5^{\prime\prime}$, ensuring that each measurement is separated by approximately half a beam.

\subsubsection{\texorpdfstring{C$^{17}$O-to-H$_2$}{C17O-to-H2} conversion factor and \texorpdfstring{$N(\mathrm{C^{18}O})/N(\mathrm{C^{17}O})$}{N(C18O)/N(C17O)} ratio} \label{Subsubsec:conversion-factor}

One key parameter for the non-LTE radiative transfer analysis is \nH{}. As the second most abundant molecule in the universe, and given the difficulties of directly observing H$_2$, CO has been historically used to trace the amount of cold molecular gas \citep{Bolatto2013}. Since CO emission is typically optically thick, observations from less abundant isotopologues are more desirable. Nevertheless, sometimes even $^{13}$CO emission is optically thick, and thus the use of rarer isotopologues, such as \CO{}, is needed. Since we have prior information about \nH{} at the PDR and dense core, we first fitted the \CO{} emission at these two specific positions to calibrate the \CO{}-to-H$_2$ conversion, and then extended it to the other $\delta x$ positions to derive an \nH{} profile.

Although the collisional rates of \CO{} with H$_2$ are available \citep{Yang2010}, we decided to model the emission using the LTE approach. This is a reasonable assumption, given that the critical density of the \CO{} $J=1-0$ transition is generally $n_\mathrm{crit} < n_{\mathrm{H_2}}$ (see Table~\ref{table:app-moldata}) for the typical gas densities (\nH{} $\sim 10^{4}-10^{5}$~cm$^{-3}$) and temperatures (\Tkin{} $\lesssim 100$~K) observed in the inner layers of the Horsehead nebula \citep{Habart2005,Pety2005}. Furthermore, we modeled the spectral lines directly, rather than the integrated intensity, to properly account for the hyperfine structure of the emission. Following the procedure described in \cite{Hernandez-Vera2024}, the line intensity ($I_{\nu}$) in LTE can be expressed in terms of $N$, \Tex{}, and $\Delta v$. However, since we only had one \CO{} transition observed with ALMA 7m+30m, it was not possible to simultaneously constrain all three parameters.

Therefore, taking advantage of its wide spectral coverage, we used WHISPER data to simultaneously fit the \CO{} $1-0$ and $2-1$ transitions. We also fitted the same transitions for C$^{18}$O to determine the C$^{18}$O/\CO{} ratio and verify the consistency of our \CO{} conversion factor with other CO isotopologues. All relevant spectroscopic constants used for the hyperfine structure fitting are provided in Table~\ref{table:app-moldata}, and the best-fit models are shown in Fig.~\ref{fig:hh-WHISPER-fits}. Each hyperfine component was modeled as an independent Gaussian with fixed velocity separations, given that they are partially blended at the $0.2$ km s$^{-1}$ resolution. Their relative intensities were set by the ratios of the upper-state degeneracies, $g_u$, and the Einstein coefficients, $A_{ul}$. From these results, we derive an average isotopologue ratio of $N(\mathrm{C^{18}O})/N(\mathrm{C^{17}O}) = 3.6 \pm 0.5$, in good agreement with the local ISM $^{18}$O/$^{17}$O element ratio \citep[3.2,][]{Wilson1994}. 

Interestingly, a \Tex{} value of $<20$~K for \CO{} and C$^{18}$O appears to be appropriate for both the PDR and dense core positions. However, given the low $n_{\mathrm{crit}}$ of their rotational transitions, LTE conditions would imply $T_{\mathrm{ex}} \approx T_{\mathrm{kin}} = 60$~K for the PDR \citep{Gerin2009}. Since the emission is barely resolved, the beam likely includes contributions from the colder, denser cloud interior rather than solely from the PDR. To test this, we repeated the analysis with \code{SpectralRadex}, fixing \nH{} to values from previous work \citep{Gerin2009}, and obtained similarly low \Tkin{} and \Tex{} values ($\sim13-20$~K), revealing beam pickup from colder material within the cloud that is likely in LTE.

Hence, with the excitation temperature fixed at $T_{\mathrm{ex}}=15$~K, we repeated the fitting procedure but now for the ALMA 7m+30m data, leaving $N$ and $\Delta v$ as free parameters. This approach allows us to determine $N$(\CO{})—needed to estimate \nH{}—while $\Delta v$ is constrained by fitting the hyperfine components of the observed transition. The best-fit models for the PDR and dense core positions are shown in Fig.~\ref{fig:hh-c17o-grid}. Defining the \CO{}-to-H$_2$ conversion factor as $X{_\mathrm{C^{17}O}}= N(\mathrm{H_2})/N(\mathrm{C^{17}O})$, we combined the $N(\mathrm{H_2})$ estimates for the PDR and dense core from \cite{Guzman2014} with our $N(\mathrm{C^{17}O})$ measurements to derive empirical conversion factors for the Horsehead nebula. The resulting values are $X_{\mathrm{C^{17}O}} = (4.0\pm0.2) \times10^{6}$ for the PDR and $X_{\mathrm{C^{17}O}} = (5.9\pm0.3) \times10^{6}$ for the dense core.

Being a less abundant isotopologue, \CO{} is not commonly used as a tracer of H$_2$ column density. As a consistency check, we used our newly derived $N(\mathrm{C^{18}O})/N(\mathrm{C^{17}O})$ ratio to compare our $X_{\mathrm{C^{17}O}}$ conversion factor with values reported in the literature. Assuming $N(^{13}\mathrm{CO})/N(\mathrm{C^{18}O}) \sim 8$ \citep{Roueff2024,Segal2024}, $N(\mathrm{C^{18}O})/N(\mathrm{C^{17}O}) \sim 3.6$ (this work), and $X_{\mathrm{C^{17}O}} = 5.0 \times10^{6}$ (average value), we determined a $^{13}\mathrm{CO}$-to-H$_2$ conversion factor of $X_{\mathrm{^{13}CO}}\sim1.7\times10^{5}$, which is somewhat lower than previous results in the Orion B molecular cloud \citep[$4.0\times10^5$,][]{Roueff2021} and other typical ISM estimates \citep[$3.8\times10^5$,][]{Bolatto2013}. Since we are observing a PDR environment, this discrepancy is likely due to more efficient selective photodissociation of \CO{} and C$^{18}$O compared to the more abundant isotopologue $^{13}$CO, as both \CO{} and C$^{18}$O become abundant only deeper into the cloud, well behind the H/H$_2$ transition.

\subsubsection{Gas density based on \texorpdfstring{C$^{17}$O}{Lg} observations}\label{Subsubsec:gas-density}

Once the $X_{\mathrm{C^{17}O}}$ conversion factor was established, we extended the same fitting procedure of \CO{} ALMA 7m+30m data to all other $\delta x$ positions at the $\delta y$ locations of the PDR and dense core, using the range shown in Fig.~\ref{fig:hh-coms-profiles}. This yielded $N(\mathrm{C^{17}O})$ profiles that we converted into \nH{} profiles. The advantage of this method is that, by using ALMA 7m+30m data, we obtain empirical gas density profiles at the same angular resolution as the other molecular line observations, thus enabling more accurate column density determinations for the organic species analyzed in this work.

Assuming a typical cloud depth along the line of sight of $\ell_{\mathrm{depth}} \simeq 0.1$~pc \citep{Habart2005,Lis2026}, and accounting for the fact that the empirical $X_{\mathrm{C^{17}O}}$ value varies between the PDR and dense core, we defined the conversion from $N$(\CO{}) to \nH{} as 

\begin{equation}
n_{\mathrm{H_2}}(\delta x) = \frac{X_{\mathrm{C^{17}O}}(\delta x)N(\mathrm{C^{17}O})(\delta x)}{ \ell_{\mathrm{depth}}},
\end{equation}

\noindent where $X_{\mathrm{C^{17}O}}(\delta x)$ is obtained by linear interpolation and extrapolation of the empirical conversion factors determined at the positions of the PDR and dense core. This approach preserves consistency with previous \nH{} determinations at these positions, while providing a smoothly varying profile between them. 

\begin{figure}[t!]
\centering
\includegraphics[width=1.0\linewidth]{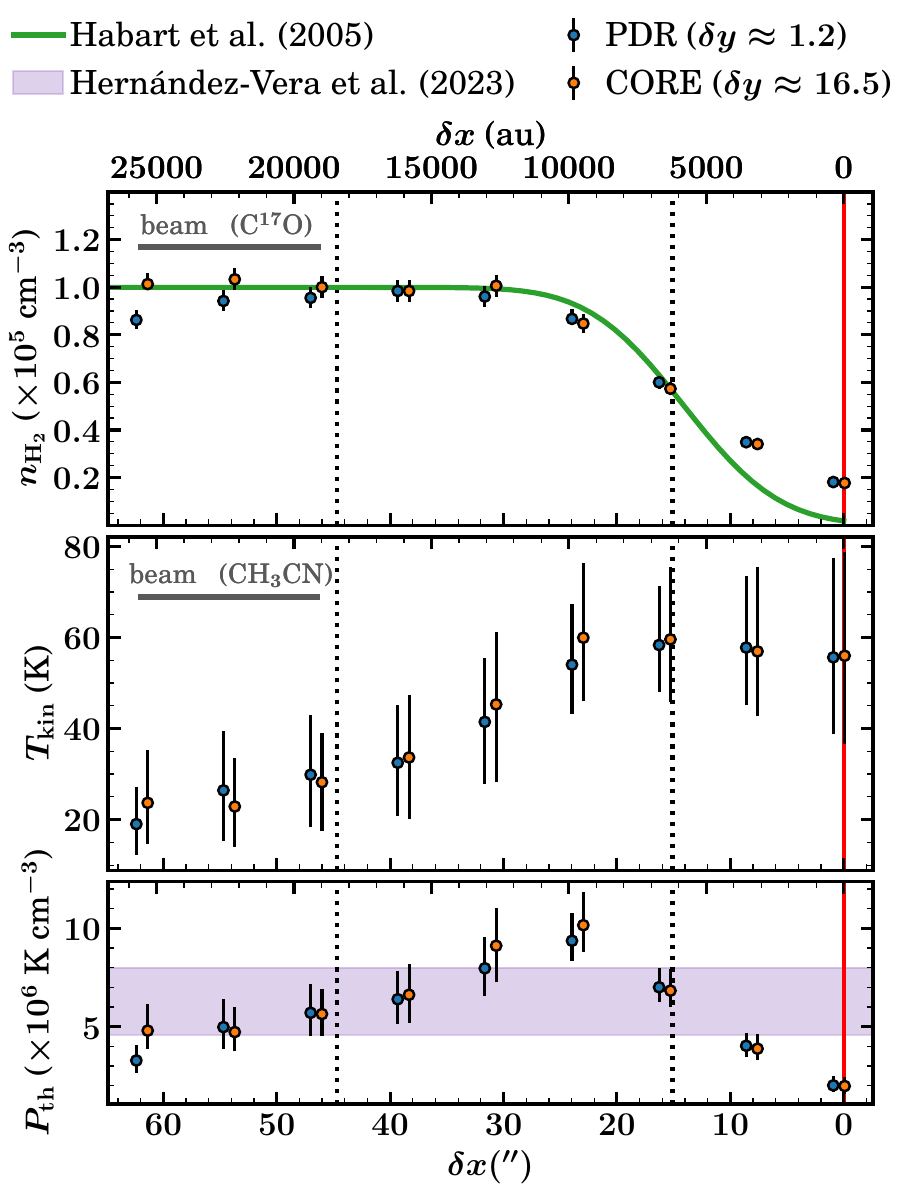}
\caption{Empirical gas density (top panel), kinetic temperature (middle panel), and thermal pressure (bottom panel) profiles determined from \CO{} and \methylcyanide{} observations of the Horsehead edge, extracted at the $\delta y$ position of the PDR (blue circles) and dense core (orange circles). The beam sizes of the respective molecular tracers from which the values were derived are represented by the horizontal gray bar in the top left corner of each panel. The red vertical line and the dotted vertical lines are the same as in the left panels of Fig.~\ref{fig:hh-coms-profiles}. For comparison, the gas density profile modeled by \cite{Habart2005} convolved with a Gaussian of 15$^{\prime\prime}$ FWHM (green line) and the thermal pressure range from \cite{Hernandez-Vera2023} (purple-shaded region) are also shown.}
\label{fig:hh-physcond-profiles}
\end{figure}

The top panel of Fig.~\ref{fig:hh-physcond-profiles} shows the resulting \nH{} profiles for the PDR and dense core vertical positions, along with the profile modeled by \cite{Habart2005} convolved with a $15^{\prime\prime}$ full width at half maximum (FWHM) Gaussian to facilitate comparison with the ALMA 7m+30m data. Both profiles confirm the presence of a steep density gradient, with typical values of $n_{\mathrm{H_2}} \sim 10^4-10^5$~cm$^{-3}$. Although lower densities in the surface layers of the cloud are also plausible \citep[$n_{\mathrm{H_2}}\lesssim 10^{3}$~cm$^{-3}$,][]{Hernandez-Vera2023}, non-LTE effects and beam pickup from colder and denser regions could still bias the inferred \CO{} column densities upward, and thus the \nH{} values derived at the edge of the cloud.

\subsubsection{Gas temperature based on \texorpdfstring{CH$_3$CN}{Lg} observations}\label{Subsubsec:gas-temperature}

\methylcyanide{} is often considered as a ``good thermometer" due to its properties of symmetric top molecule \citep[e.g., ][]{Guesten1985}. Considering that there cannot be radiative transitions between different $K$-ladders, the energy levels with different $K$ values are populated by collisions. Therefore, the relative intensities between transitions with the same $J$ but different $K$ values are sensitive to \Tkin{}.

\begin{figure*}[t!]
\centering
\includegraphics[width=1.0\linewidth]{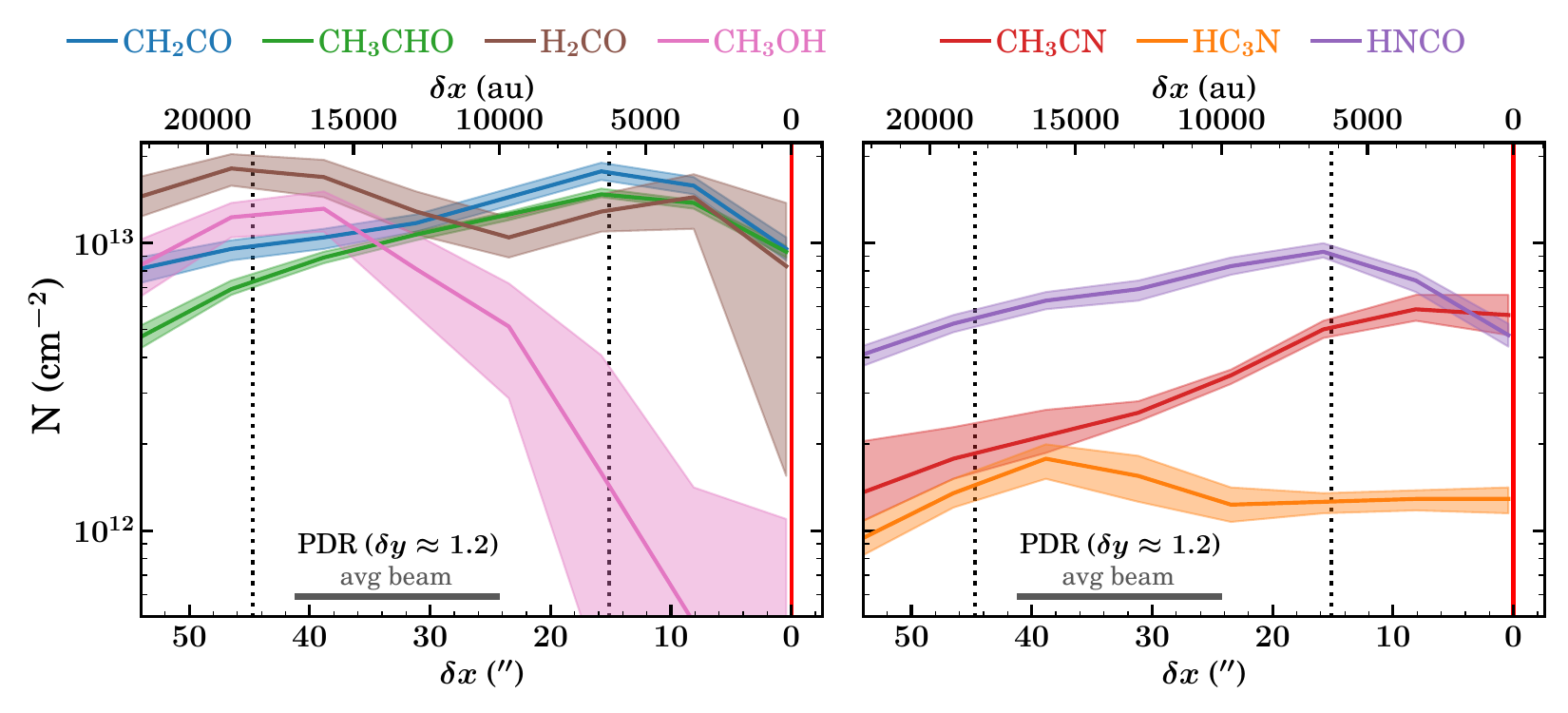}
\caption{Column density profiles derived at the $\delta y$ position of the PDR from the radiative transfer modeling of the O-bearing (left panel) and N-bearing (right panel) molecules analyzed in this work. For each color the solid line depicts the best-fit values, whose uncertainties are represented by the shaded areas of the same color. The red vertical line, the dotted vertical lines, and the horizontal gray bar are the same as in the left panels of Fig.~\ref{fig:hh-coms-profiles}. The same profiles, but at the  $\delta y$ position of the dense core, are shown in Fig.~\ref{fig:hh-logN-profiles-CORE}.}
\label{fig:hh-logN-profiles}
\end{figure*}

As described at the beginning of Sect.~\ref{Subsec:radiative-transfer}, we used \code{SpectralRadex} to model the integrated line intensity of the three \methylcyanide{} $J = 6-5$ $(K=0,1,\mathrm{and}\:2)$ transitions simultaneously. In this case, we decided to employ the non-LTE approach since \cite{Gratier2013} demonstrated that \methylcyanide{} emission at the Horsehead PDR is subthermally excited. The collisional rates were taken from \cite{BenKhalifa2023}, and we assumed a ratio of $\mathrm{E/A} = 1$ between the two noninteracting torsional substates. Using the \nH{} values derived from \CO{} as input, the modeled \methylcyanide{} intensities only depend on \Tkin{} and $N(\mathrm{CH_3CN})$ as free parameters. The best-fit integrated intensity models are shown in Fig.~\ref{fig:hh-ch3cn-fit}. The resulting \Tkin{} profiles for the PDR and dense core vertical positions are shown in the middle panel of Fig.~\ref{fig:hh-physcond-profiles}, whereas the $N(\mathrm{CH_3CN})$ profiles are shown in Fig.~\ref{fig:hh-logN-profiles} and Fig.~\ref{fig:hh-logN-profiles-CORE}.

We have obtained temperature values of $T_{\mathrm{kin}} = 58^{+13}_{-10}$~K for the PDR position ($\delta x \approx 15\farcs1,\:\delta y \approx 1\farcs2$) and $T_{\mathrm{kin}} = 28^{+11}_{-10}$~K for the dense core position ($\delta x \approx 44\farcs7,\:\delta y \approx 16\farcs5$), which is in good agreement with previous estimates using \methylcyanide{} \citep{Gratier2013} and other molecular tracers \citep{Pety2007,Gerin2009,Guzman2013}. Regarding the other profile positions, we have empirically demonstrated the presence of a temperature gradient, as earlier pointed out by observations \citep{Pety2005} and models \citep{Goicoechea2009a,LeGal2017,LeGal2019}. In this case, the temperature derived from \methylcyanide{} at the PDR is significantly higher than that estimated from \CO{} using WHISPER observations. Although the beam sizes are not very different in the $\delta x$ direction, the spatial distribution of \CO{} is much more extended than that of the \methylcyanide{} filament, which shows almost no counterpart in the dense core and likely minimizes beam-pickup effects from colder interior regions.

However, we did not recover the high temperatures ($T_{\mathrm{kin}} \gtrsim 100$~K) expected at the very edge of the cloud \citep{Zannese2025b}, in part because the collisional rates used were tabulated only up to $100$~K. Still, the posterior probability distributions showed no indication of a preference for higher temperatures. This discrepancy likely stems from the limited angular resolution and sensitivity of the observations, which may prevent detection of the outermost regions of the cloud, characterized by lower column densities and, consequently, lower surface brightness. 

\subsubsection{Gas thermal pressure}\label{Subsubsec:gas-pressure}

Having derived empirical profiles for the gas density and kinetic temperature, we computed the corresponding thermal pressure, defined as $P_{\mathrm{th}} = n_{\mathrm{H}} T_{\mathrm{kin}}$, where $n_{\mathrm{H}}=n(\mathrm{H})+2n(\mathrm{H_2})\sim 2n_{\mathrm{H_2}}$ corresponds to the total hydrogen nuclei density. The resulting $P_{\mathrm{th}}$ profiles for the vertical positions of the PDR and dense core are shown in the bottom panel of Fig.~\ref{fig:hh-physcond-profiles}, together with the range of values determined from high angular resolution observations by \citet{Hernandez-Vera2023}.\footnote{In \cite{Hernandez-Vera2023} the factor of 2 in the definition of $n_{\mathrm{H}} \sim 2n_{\mathrm{H_2}}$ is missing. As a result, and for consistency with previous work, the range of values used for comparison here is $P_{\mathrm{th}} = (4.6$–$8.0)\times10^{6}$~K~cm$^{-3}$.}

These results corroborate the presence of high thermal pressures at the Horsehead edge, exceeding those measured in the adjacent IC 434 \HII{} region \citep{Bally2018}, in agreement with a photoevaporative compression scenario. Moreover, to first order, the Horsehead PDR appears to maintain an isobaric structure: although some variations are present, most values remain, within uncertainties, consistent with the range reported by \citet{Hernandez-Vera2023}. These findings add compelling new observational support to previous evidence that PDRs can often be represented as isobaric regimes with large thermal pressures regulated by the intensity of the incident UV radiation field \citep{Goicoechea2016,Joblin2018,Wu2018,Bron2018,Maillard2021,Hernandez-Vera2023,Clark2025}.

\subsubsection{Column density profiles}\label{Subsubsec:column-densities}

The integrated line emission of the remaining molecules was modeled to obtain column density profiles and further investigate the formation pathways of the different chemical species. Although it was mentioned that complex molecules in the Horsehead nebula are subthermally excited (i.e., $n_{\mathrm{crit}}<n_{\mathrm{H_2}}$, see Table~\ref{table:app-moldata}), we cannot use the RADEX non-LTE approach for all the species presented in this work since not all have collisional rates available in the literature. 

Thus, for those species with available rates, such as \formaldehyde{} \citep{Wiesenfeld2013}, \methanol{} \citep{Rabli2010}, \isocyanicacid{} \citep{Sahnoun2018}, and \cyanoacetylene{} \citep{Faure2016}, we performed a similar procedure as \methylcyanide{} but using the \nH{} and \Tkin{} profiles derived in the previous sections as input. Instead, for \ketene{} and \acetaldehyde{}, we used the rotational diagram approach by assuming a Boltzmann distribution of the rotational levels characterized by a single \Tex{} lower than \Tkin{}, consistent with a subthermal excitation regime. Under this assumption, the column density in the upper state of a given transition ($N_u$) is given by

\begin{equation}
    \label{eq:Boltzmann}
    N_u = \frac{N}{Q(T_{\mathrm{ex}})}g_ue^{-E_u/kT_{\mathrm{ex}}},
\end{equation}

\noindent where $N$ is the total column density, $Q$ is the partition function of the molecule taken from the JPL database \citep{Pickett1998}, and $E_u$ with $g_u$ are the energy and the degeneracy associated with the upper-state level of the corresponding transition, respectively (see Table~\ref{table:app-moldata}). On the other hand, assuming optically thin emission, $N_u$ is related to the integrated line intensity ($I = \int I_\nu d\nu$, in K~km~s$^{-1}$) through the equation

\begin{equation}
    \label{eq:Int_intensity}
    I = \frac{hc^3A_{ul}}{8\pi k\nu^2}N_u,
\end{equation}

\noindent where $\nu$ is the line frequency and $A_{ul}$ is the Einstein coefficient (see Table~\ref{table:app-moldata}). By putting Eq. (\ref{eq:Boltzmann}) into Eq. (\ref{eq:Int_intensity}), the integrated line intensity can therefore be written as a function of $N$ and $T_{\mathrm{ex}}$. Similar to the \CO{} case, we fixed the excitation temperature at $T_{\mathrm{ex}} = 15$~K for \ketene{} and $T_{\mathrm{ex}} = 8$~K for \acetaldehyde{}, based on previous work \citep{Guzman2014}. 

Hence, either using the non-LTE approach implemented by \code{SpectralRadex} or the rotational diagram approach through Eqs. (\ref{eq:Boltzmann}) and (\ref{eq:Int_intensity}), the only free parameter to fit is $N$. The fitting strategy was essentially the same as for \methylcyanide{}, which is illustrated in Appendix~\ref{subApp:CH3CN}. It is worth mentioning that, similar to the case of \methylcyanide{}, we assumed $\mathrm{E/A} = 1$ for \methanol{} and \acetaldehyde{} and an ortho-to-para ratio of $\mathrm{OPR} = 3$ for \formaldehyde{} and \ketene{} since we do not have any information about these symmetry ratios and nuclear spin forms in positions other than the PDR and the dense core. In these cases, the total column density corresponds to the sum of the different symmetries (E+A) or nuclear spin forms (ortho+para).

Figure~\ref{fig:hh-logN-profiles} shows the derived column density profiles for the PDR vertical position, while the corresponding profiles for the dense core vertical position are presented in Fig.~\ref{fig:hh-logN-profiles-CORE}. Since the profiles at both positions are very similar, we   refer only to Fig.~\ref{fig:hh-logN-profiles} in the subsequent analysis, with the understanding that the same conclusions apply to the dense core vertical position. The associated uncertainties correspond to the statistical propagation of observational errors, without including systematic uncertainties on the fixed parameters, as this would require exploring a multidimensional parameter space in density and temperature, which is beyond the scope of this work. We checked the consistency of our results by comparing them with previous efforts. Since this is the first time that the column density of these molecules is spatially imaged, the only benchmarks that we have are the values derived at the PDR and dense core positions from prior single-dish observations, except for HNCO for which there are no results reported in the literature.

For the molecules observed with the PdBI, our results are consistent within the uncertainties with the values determined by \cite{Guzman2013} using the IRAM 30m spectra only, confirming that the emission of \formaldehyde{} and \methanol{} is only marginally diluted. Instead, for the combined ALMA 7m+30m maps, most of our column densities are higher than previous results from the WHISPER line survey \citep{Gratier2013,Guzman2014}, specifically by a factor of $\sim3-4$ in the dense core and a factor of $\sim 3-6$ in the PDR. Beam dilution effects can explain the above since, from the maps shown in Fig.~\ref{fig:hh-coms-grid}, we can see that we begin to resolve filamentary and clumpy structures with angular widths on the order of $\theta_{\mathrm{source}}\sim 10-15^{\prime\prime}$ that probably were diluted by the large beam size reached by WHISPER \citep[$\theta_{\mathrm{beam}}\gtrsim 27^{\prime\prime}$,][]{Guzman2014}.

For \methylcyanide{}, at the PDR position we found a column density $\sim0.5$ times lower than that reported by \cite{Gratier2013} using WHISPER data. However, \cite{Gratier2013} already accounted for beam dilution in their radiative transfer modeling, which explains why we find a similar column density with the new higher angular resolution observations. At the dense core, in contrast, we find a higher \methylcyanide{} column density (by a factor of $\sim3$) compared to \cite{Gratier2013}. This discrepancy is likely due to the choice of collisional rates: \cite{Gratier2013} adopted the rates from \cite{Green1986}, whereas \cite{BenKhalifa2023} recently showed that those rates can be severely underestimated for some transitions, especially at low temperatures ($T_{\mathrm{kin}} = 20$~K) such as those found in the dense core. When we repeated our analysis using the older rates from \cite{Green1986}, our derived $N$ values matched those reported by \cite{Gratier2013}. In the PDR, on the other hand, the choice of collisional coefficients has a less significant effect.

Concerning the shape of the $N$ profiles, we find that \isocyanicacid{}, \methylcyanide{}, \ketene{}, and \acetaldehyde{} present higher column densities at the edge of the Horsehead nebula, with their peak near the PDR position. This suggests that their formation is enhanced in the presence of FUV radiation, consistent with the modern view of PDRs in which FUV photons can promote the production of certain molecules \citep[e.g.,][]{Goicoechea2025}. In contrast, we find that \methanol{} and \cyanoacetylene{} present higher column densities in deeper layers of the Horsehead nebula, with their peak near the dense core. These molecules are probably formed (or survive) more efficiently or readily desorbed into the gas-phase when shielded from FUV radiation. Lastly, \formaldehyde{} stands out for its high abundance in both FUV-exposed and FUV-shielded layers of the Horsehead nebula, indicating that it can form efficiently under a range of physical conditions.

\section{Discussion}\label{Sec:discussion}

\subsection{Chemistry of O-bearing species}\label{Subsec:O-discussion}

Figure~\ref{fig:hh-logN-profiles} reveals clear differences in the column density distributions of molecular species with similar elemental compositions, likely reflecting distinct underlying chemical processes. In the case of O-bearing species, \formaldehyde{} and \methanol{} have their peak closer to the dense core position, whereas \ketene{} and \acetaldehyde{} exhibit larger values near the PDR. This may seem surprising, considering that both molecular pairs are thought to be efficiently formed on dust grain surfaces through CO hydrogenation \citep[see][and references therein]{Herbst2009}. However, the observed chemical differentiation may arise from the presence of an additional C atom that distinguishes the two groups of molecules.

Recent estimates of the binding energy of atomic C onto grain surfaces are reshaping the classical view proposed by early PDR models \citep[e.g.,][]{Hollenbach2009}. These models typically assumed a low binding energy \citep[$E_{\mathrm{b}} \sim 800$~K,][]{Hasegawa1993}, implying that atomic C would freeze out onto grains only at very low temperatures ($T_{\mathrm{kin}} \lesssim 20$~K). However, more recent quantum chemical calculations suggest much higher binding energies \citep[$E_{\mathrm{b}} > 10^{4}$~K,][]{Shimonishi2018,Duflot2021,Minissale2022}, indicating that atomic C can deplete at significantly higher temperatures onto amorphous water ices, as those expected to coat grains within the interior of the Horsehead nebula \citep{Schirmer2020,Abergel2024}.

This is supported by the fact that laboratory experiments have demonstrated that the surface diffusion of C atoms can be a considerable alternative to increase the complexity of organic molecules in translucent clouds \citep{Tsuge2023}. Certainly, previous studies have suggested that the formation of \ketene{} and \acetaldehyde{} can be initiated by the barrierless reaction C + CO $\xrightarrow{}$ C$_2$O \citep{Maity2014,Fedoseev2022,Borshcheva2025}, avoiding HCO as an intermediate product, which is the common ancestor with \formaldehyde{} and \methanol{}. Additionally, the grain surface chemistry of \ketene{} and \acetaldehyde{} can be favored by reactions involving CH$_2$ \citep{Chen2025} and C$_2$H$_2$ \citep{Chuang2020}, which can be also produced by C diffusive reactions \citep{Tsuge2023,Ferrero2024}.

Near the PDR position, the presence of FUV radiation leads to the photodissociation of CO, resulting in the coexistence of CO, atomic C, and C$^+$ at the illuminated edge of the Horsehead nebula \citep{Philipp2006,Bally2018,Hernandez-Vera2023}. Atomic C can then be physisorbed onto H$_2$O-rich ices on shallow binding sites, allowing its efficient diffusion for dust temperatures above $\sim$22~K \citep{Tsuge2023}, consistent with the conditions expected at the Horsehead PDR \citep[$T_{\mathrm{dust}} \sim 30$~K,][]{Goicoechea2009b}. In contrast, in the dense core where dust temperatures are lower \citep[$T_{\mathrm{dust}} \sim 20$~K,][]{Goicoechea2009b}, CO freeze-out becomes more relevant and atomic C diffusion is less efficient. Under these conditions, CO hydrogenation likely dominates, enhancing the formation of \formaldehyde{} and \methanol{}.

Interestingly, although \formaldehyde{} is more abundant near the dense core, it also shows a significant contribution from the PDR, comparable to that of \ketene{} and \acetaldehyde{}. This suggests that atomic C may also play a key role in the grain-surface formation of \formaldehyde{}. In fact, an alternative to the traditional CO hydrogenation pathway involves the surface reaction C + H$_2$O $\rightarrow$ H$_2$CO, recently supported by both computational studies \citep{Molpeceres2021} and laboratory experiments \citep{Potapov2021}. Consistently, \cite{Tsuge2023} found that a considerable fraction of the atomic C physisorbed onto H$_2$O-rich ices reacts rapidly and forms \formaldehyde{}.

An increased production of \formaldehyde{} on grains enhances the cold formation pathways of \ketene{} and \acetaldehyde{}, but also \methanol{} \citep{Potapov2024}. However, under FUV irradiation, \methanol{} tends to fragment into smaller radicals, which can subsequently lead to the formation of \ketene{} and \acetaldehyde{} \citep{Yocum2021}, or even regenerate \formaldehyde{} \citep{Bertin2016}. Furthermore, while \methanol{} is predominantly formed via surface reactions on dust grains \citep{Watanabe2002,Fuchs2009}, \ketene{}, \acetaldehyde{}, and \formaldehyde{} also follow viable gas-phase formation pathways \citep{Ruiterkamp2007,Vazart2020,Ramal-Olmedo2021}, as shown by recent detections of these species in diffuse gas lacking ice mantles \citep{Gerin2025}.

These pathways typically involve reactions between atomic O and hydrocarbon radicals—specifically, C$_2$H$_3$, C$_2$H$_5$, and CH$_3$ for \ketene{}, \acetaldehyde{}, and \formaldehyde{}, respectively—highlighting another possible chemical link among the three species. Notably, atomic O has been detected at the edge of the Horsehead nebula \citep{Goicoechea2009b}, and the hydrocarbon radicals may originate from the photodissociation of closed-shell hydrocarbons like C$_2$H$_6$ and CH$_4$ \citep{Heays2017}, which themselves are efficient products of grain-surface C chemistry \citep{Kobayashi2017,Qasim2020,Tsuge2024}. This would reflect a form of top-down hydrocarbon chemistry, complementing that previously proposed for PAHs at the edge of the Horsehead nebula \citep{Pety2005,Guzman2015}. 

Overall, whether O-bearing molecules form on dust grains or in the gas phase, the presence of atomic C under FUV-illuminated conditions appears to play a key role in driving their chemical complexity at relatively low temperatures ($T_{\mathrm{kin}} \lesssim 100$~K). In the dense core, by contrast, CO hydrogenation leading to \formaldehyde{} and \methanol{} becomes more important, since atomic C is less abundant and its diffusion on grain surfaces is less efficient. The UV-shielded conditions also limit the fragmentation of \methanol{}, thereby influencing the formation of \ketene{} and \acetaldehyde{}. Notably, the prominent detection of \methanol{}—and O-bearing species in general—in the gas phase of the dense core may serve as direct evidence of chemical desorption, as previously suggested by \citet{LeGal2017}.

\subsection{Chemistry of N-bearing species}\label{Subsec:N-discussion}

Figure~\ref{fig:hh-logN-profiles} also shows that N-bearing species display notably different column density distributions. While \methylcyanide{} shows a steep profile peaking near the PDR position, \cyanoacetylene{} presents a more gradual increase toward the dense core. In contrast, \isocyanicacid{} exhibits similarities with the O-bearing molecules, as previously indicated by its integrated intensity distribution (see Fig.~\ref{fig:hh-coms-profiles}). In line with our interpretation for the O-bearing species, we propose that the chemistry of N-bearing molecules in FUV-irradiated regions is likewise influenced by atomic C chemistry.

According to the literature, \methylcyanide{} can be initiated in the gas phase through the radiative association reaction CH$_3^+$ + HCN $\rightarrow$ CH$_3$CNH$^+$ + $h\nu$, followed by dissociative recombination with electrons \citep{Giani2023}. Nevertheless, its enhanced column density near the PDR may indicate a contribution from dust grain surface chemistry, as previously suggested by \cite{Gratier2013} and supported by experimental evidence showing that \methylcyanide{} undergoes more efficient photodesorption than molecules such as \methanol{} \citep{Basalgete2021}. One proposed grain-surface formation pathway involves the reaction between CH$_3$ and CN radicals to form \methylcyanide{} \citep{Garrod2008,Enrique-Romero2025}, with both radicals potentially originating from the diffusion of atomic C or the photodissociation of larger species. 

Another possible route of \methylcyanide{} on grains is through the hydrogenation of C$_2$N \citep{Walsh2014}. This molecule is thought to form primarily in the gas-phase via the reaction C$_2$H + N $\rightarrow$ H + C$_2$N, followed by freeze-out onto dust grains \citep{Loomis2018}. However, there are grain-surface reactions like C + CN or C$_2$ + N $\rightarrow$ C$_2$N listed as potential pathways in the Kinetic Database for Astrochemistry \citep[KIDA;][]{Wakelam2012,Wakelam2024} that have yet to be experimentally confirmed. Additionally, \cite{Canta2023} probed that the formation of \methylcyanide{} is also possible due to the photoprocessing of ice mixtures composed by NH$_3$ and hydrocarbons of the form C$_2$H$_{\mathrm{x}}$, where the latter are possibly produced by atomic C on dust grains.

Despite the above, gas-phase chemistry may still play a significant role, as predicted by PDR models \citep{LeGal2019}. CH$_3$$^+$, a key gas-phase precursor of \methylcyanide{}, can form efficiently in FUV-irradiated gas where C$^+$ is abundant and vibrationally excited H$^*_2$ ($v \geq 1$) is present \citep{Berne2023,Goicoechea2025}, such as in PDRs \citep{Zannese2025}. These same conditions also promote the efficient formation of HCO$^+$ \citep{Berne2024}. Notably, all these requirements (C$^+$, H$^*_2$, and HCO$^+$) are met in the Horsehead PDR, near the bright \methylcyanide{} filament \citep{Bally2018,Hernandez-Vera2023,Abergel2024}. In contrast, in the FUV-shielded dense core, this mechanism is likely not relevant.

In the case of \cyanoacetylene{}, it is generally accepted that this molecule lacks efficient grain-surface formation pathways \citep{Bergner2017,Bergner2018}. This may account for the absence of a filamentary emission structure at the PDR, in contrast to most of the species discussed in this work, which likely receive significant contributions from dust grain chemistry. In the gas phase, the formation of \cyanoacetylene{} is primarily driven by hydrocarbon reactions such as C$_2$H$_2$ + CN and C$_2$H + HCN $\rightarrow$ HC$_3$N + H \citep{Fukuzawa1997,Taniguchi2024}. A less dominant, yet nonnegligible, route involves C$_3$H$_2$ + N $\rightarrow$ HC$_3$N \citep{Murillo2022}. In the Horsehead nebula, these hydrocarbon precursors (e.g., C$_2$H, C$_2$H$_2$, and C$_3$H$_2$) are believed to originate mainly from gas-phase reactions initiated by C$^+$ and CH, along with PAH photochemistry \citep{Guzman2015}. However, an alternative origin linked to products of grain-surface C chemistry, such as CH$_4$ \citep{Lamberts2022,Tsuge2024}, may also contribute \citep[e.g.,][]{Hassel2008,Sakai2013,Loison2017}.

PdBI maps of C$_2$H and $c$-C$_3$H$_2$ provide additional evidence that these species are chemically linked to \cyanoacetylene{}. \citet{Guzman2015} noted that the emission of these hydrocarbons follows the same two-filament structure recently revealed by high angular resolution observations of HCO$^+$, located in the vicinity of the PDR and dense core \citep{Hernandez-Vera2023}. At the PDR, \cyanoacetylene{} does not display a noticeable filament, likely because it is more efficiently destroyed than formed—consistent with its rapid photodissociation relative to other N-bearing species such as \methylcyanide{} \citep{LeGal2019}, as well as the lack of efficient grain-surface formation pathways. In contrast, within the filament near the dense core, all three molecules trace the same emitting region, although C$_2$H appears to be depleted where $c$-C$_3$H$_2$ and \cyanoacetylene{} are brighter. Moreover, at the exact position of  the dense core—where HCO$^+$ and DCO$^+$ reach their emission peaks \citep{Pety2007,Hernandez-Vera2023}—\cyanoacetylene{} shows a slight depletion, in line with its predicted rapid destruction via ion–molecule reactions \citep{Taniguchi2024}.

Lastly, although \isocyanicacid{} is composed by nitrogen and oxygen, we decided to consider it as a N-bearing molecule due to the key role of atomic N during its formation under FUV-irradiated conditions. Despite its relative simplicity, the formation pathways of \isocyanicacid{} remain poorly constrained. Current models suggest a combined contribution from both gas-phase and grain-surface chemistry, though disentangling their respective roles is challenging \citep{Tideswell2010,Quan2010}. In both cases, the radical NCO is considered a key intermediate \citep{Marcelino2010}. In the gas phase, \isocyanicacid{} can be formed via protonation of NCO by proton donors like H$_3^+$, H$_3$O$^+$, and HCO$^+$, followed by further ion–molecule reactions and eventual dissociative recombination \citep{VelillaPrieto2015,Cernicharo2024}. On grain surfaces, hydrogenation of NCO has been proposed as an efficient alternative as well \citep{Quan2010,Belloche2017}.

However, the chemistry of NCO itself is also intricate. Its formation is typically modeled in the gas phase through reactions such as N + HCO → NCO + H \citep{Quan2010,Cernicharo2024}, CN + O$_2$ → NCO + O \citep{Marcelino2018,Cernicharo2024}, and CN + OH → NCO + H \citep{VelillaPrieto2015}. Once produced, NCO may accrete onto dust grains \citep{Quan2010}, although direct formation on icy mantles is also possible via reactions between CO and atomic N \citep{Fedoseev2015,Mignon2017}, or through pathways involving CO photodissociation products such as O + CN and C + NO \citep{Belloche2017}. Notably, the latter reaction may benefit from efficient surface diffusion of atomic C.

\begin{figure*}[t!]
\centering
\includegraphics[width=1.0\linewidth]{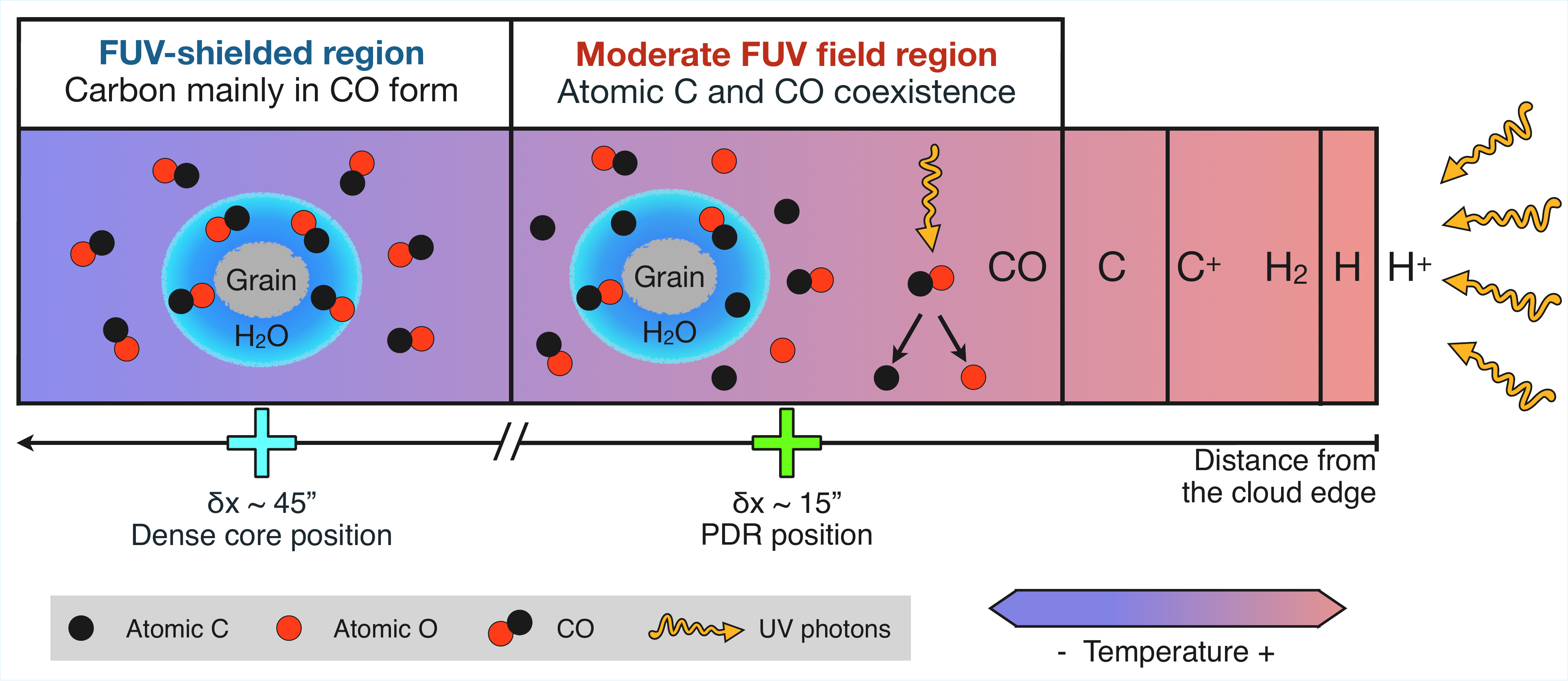}
\caption{Schematic summary of the proposed scenario at the edge of the Horsehead nebula. The two crosses indicate the PDR and dense core positions, as in Fig.~\ref{fig:hh-coms-grid}. At the PDR, the moderate FUV field enables the coexistence of C and CO, together with a sufficiently high $T_\mathrm{dust}$ to allow efficient diffusion of atomic C on grain surfaces. These conditions may favor the formation of \formaldehyde{}, \ketene{}, \acetaldehyde{}, \isocyanicacid{}, and \methylcyanide{}, likely through a combination of gas-phase and grain-surface pathways.
Interestingly, for the two species with the lowest PDR abundances, \methanol{} and \cyanoacetylene{}, previous studies suggested ineffective photodesorption and inefficient grain-surface formation, respectively. In the case of the dense core, the FUV-shielded conditions limit CO photodissociation, and the low $T_\mathrm{dust}$ makes the diffusion of atomic C on grains inefficient. These conditions may favor the grain-surface formation of \formaldehyde{} and \methanol{}, and also contribute to the prevalence of \cyanoacetylene{} in the gas phase as it is less efficiently photodissociated.}
\label{fig:hh-summary-fig}
\end{figure*}

Another important formation route for \isocyanicacid{} involves the simultaneous hydrogenation and UV processing of gas-phase NO after its accretion onto CO, H$_2$CO, and CH$_3$OH-rich ices \citep{Fedoseev2016}. This mechanism is particularly relevant given the abundant detection of NO at the edge of the Horsehead nebula (WHISPER collaboration, private communication). Laboratory experiments also suggest that the irradiation of ice mixtures containing other N-bearing species (e.g., N$_2$, NH$_3$, and HCN) may contribute to \isocyanicacid{} formation through complementary pathways \citep{Jimenez-Escobar2014,Martin-Domenech2020,Martin-Domenech2024}. Such processes may help explain the elevated column densities observed near the PDR position. On the other hand, hydrogenation products of atomic N, such as NH and NH$_2$, can efficiently react with CO on grain surfaces to form \isocyanicacid{} \citep{Fedoseev2015,Belloche2017}, in a manner analogous to the surface chemistry proposed for O-bearing species. This could account for the relatively high column densities of \isocyanicacid{} near the dense core, compared to other N-bearing molecules.

Similar to O-bearing species, these findings suggest that the chemistry of N-bearing molecules is closely linked to that of atomic C under FUV-irradiated conditions, both in the gas phase and on dust grain surfaces. Nonetheless, atomic N and its derivative species may also play a significant role, particularly in the case of \isocyanicacid{}, which contains fewer C atoms in its structure. To achieve a more comprehensive understanding of the formation and evolution of HNCO and N-bearing molecules in general, further dedicated studies targeting smaller N-bearing species (e.g., HCN, HNC, HNO, NCO, CN, and NO) in combination with detailed chemical modeling, are likely necessary.

Figure~\ref{fig:hh-summary-fig} presents a schematic overview of the physicochemical processes proposed in this work. Despite the experimental evidence and observational signposts that atomic C may contribute to the formation of COMs, modeling its surface chemistry remains challenging. A recent study by \cite{Lis2026}, using the Meudon PDR code \citep{LePetit2006} wrapped in a locally circular 2D geometry, found difficulties in reproducing the observed water vapor emission in the Horsehead nebula, suggesting that current models may be missing key grain-surface or photochemical processes. This highlights the need for further refinement of surface chemistry treatments, which appear to be fundamental for accurately modeling the abundances of COMs.

\subsection{Implications in other UV-irradiated environments}\label{Subsec:UVenvs-discussion}

The insights gained from the Horsehead nebula can inform our understanding of the formation of COMs in other UV-irradiated environments. Just as the Horsehead PDR serves as a benchmark for regions where molecular gas is exposed to moderate FUV radiation, the Orion Bar PDR has traditionally been regarded as a prototypical case for strongly FUV-irradiated molecular gas \citep{Tielens1993,Hogerheijde1995,Goicoechea2016}.

As a consequence of the higher FUV flux in the Orion Bar PDR \citep[$G_0 > 10^4$,][]{Marconi1998,Peeters2024}, the degree of photodissociation is expected to be significantly greater than in the Horsehead PDR. Furthermore, dust grains in the Orion Bar are warmer \citep[$T_{\mathrm{dust}} \simeq 40$--$80$~K,][]{Arab2012}, which inhibits the formation of substantial ice mantles and thereby reduces the efficiency of grain surface chemistry. We have found that most of the molecules analyzed in this work are less abundant in the gas phase of the Orion Bar PDR \citep{Cuadrado2017} compared to the Horsehead PDR, with the remarkable exception of \formaldehyde{} and \methanol{} (see Table~\ref{table:abundances}).

Given that \methanol{} is predominantly formed on grains—with \formaldehyde{} as one of its main precursors—a nonnegligible grain-surface contribution deep inside the Orion Bar cannot be ruled out. \citet{Cuadrado2017} showed that pure gas-phase models fail to reproduce the observed abundances of \formaldehyde{} and \methanol{}, suggesting that the formation of organics on warm grains with limited ice mantles may still be feasible. Nevertheless, given the high abundance of C at the edge of the Orion Bar \citep{Goicoechea2025} and the elevated dust temperatures, an efficient surface diffusion of atomic C on grains should play a significant role, but apparently is not the case. Hence, in contrast to the Horsehead PDR, where C can be sufficiently abundant on grains to compete with CO hydrogenation, the Orion Bar PDR appears to process ices dominated by CO rather than C, resembling those expected in the Horsehead dense core. This is, however, unlikely to occur in situ in the Bar owing to the high temperatures—though it may operate in dense cloud interiors, as evidenced by spatially resolved \methanol{} observations \citep{Leurini2010}.

A possible explanation to these contrasting behaviors in surface chemistry products may lie in PDR dynamics. \citet{Cuadrado2017} already speculated that, in the Orion Bar, the advection of material from the cold, UV-shielded interior of the molecular cloud to the warmer PDR edge is plausible in the presence of considerable dynamical effects. Supporting this, \citet{Goicoechea2016} demonstrated that dynamical processes can significantly shape the cloud structure in the Orion Bar. In the case of the Horsehead, although dynamical effects are expected to be even stronger than in high-excitation PDRs \citep{Maillard2021}, the present results, together with the absence of photoablative molecular gas flows and the lower compression factors \citep{Hernandez-Vera2023}, could suggest a less prominent dynamical influence compared to the Orion Bar, at least concerning COMs formation. Nevertheless, higher angular resolution observations are required to confidently assess this scenario.

Furthermore, a different ice composition affect not only grain-surface products but also the gas-phase processes. In the Orion Bar PDR, the intense UV radiation field likely results in nearly bare grains, depleted of H$_2$O and especially CO \citep{Goicoechea2025}. This leads to an O-rich gas-phase environment at the PDR position, implying $\mathrm{C/O}<1$. Despite this, the higher gas temperatures and enhanced abundances of H$_2^{*}$, CH$^{+}$, CH$_3^{+}$, and other hydrocarbons drive an active hydrocarbon chemistry \citep{Cuadrado2015,Goicoechea2025,Zannese2025}. Conversely, the lower dust temperatures in the Horsehead PDR allow for the formation of H$_2$O ice—and to some extent CO ice—even within the UV-irradiated layers, favoring higher gas-phase C/O ratios \citep{LeGal2019}. Considering the interplay between gas-phase and grain-surface chemistry previously discussed, these fundamental differences in ice and gas composition likely contribute to the observed variations in COM abundances. 

Our spatially resolved analysis of complex organics in the Horsehead nebula can also have important implications for protoplanetary disks. Driven primarily by UV radiation and X-rays from the central star, the chemical composition observed in the cold outer regions of nearby disks qualitatively resembles that expected for classical PDRs \citep{Oberg2021,Oberg2023}. Rich in nitriles and hydrocarbons, yet notably deficient in large O-bearing organics, these disk regions are typically characterized by a high C/O ratio, a scenario similarly proposed for the Horsehead PDR. Furthermore, recent ALMA observations have uncovered extended [C{ \small I}] emission in protoplanetary disks \citep{Booth2023,Law2023,Urbina2024}, possibly indicating that atomic C could have a relevant yet largely unexplored chemical role within these environments.

In this work, we find that \methanol{} is not the dominant O-bearing species in the gas phase under moderate FUV irradiation, which may help explain its elusive detection in cold Class II disks \citep[e.g.,][]{Carney2019,Loomis2020,Yamato2024}. In contrast, \formaldehyde{} is abundantly detected in disks \citep[e.g.,][]{Pegues2020,Guzman2021}, yet its potential formation via grain-surface reactions involving atomic C remains unexplored. If efficient, this pathway would increase \methanol{} production, whose photofragmentation might drive further chemical complexity, as observed in the Horsehead PDR. Similarly, N-bearing species like \methylcyanide{} and \cyanoacetylene{} are also common in disks \citep{Bergner2018,Ilee2021}, with their formation likely tied to UV-driven chemistry \citep{Calahan2023}.

Although dust grains in protoplanetary disks typically settle toward the cold and UV-shielded midplane \citep{Blum2008}, turbulent mixing can transport icy material into more irradiated layers \citep{Oberg2015,BergnerCiesla2021,Ligterink2024}. Models including this process predict notable effects for species like \ketene{}, \acetaldehyde{}, and \isocyanicacid{} \citep{Semenov2011}, which—despite their prebiotic relevance—are rarely targeted in disk studies. Recent observations by \citet{Yamato2024} also reveal abundant CH$_3$OCH$_3$ in the cold disk MWC 480 due to in situ ice processing. Altogether, these results suggest that PDR-like chemistry provides valuable context for disk studies, motivating the search for tracers beyond \methanol{}.

\section{Summary and Conclusions} \label{Sec:summary-conclusions}

We have investigated the spatial distribution of simple and complex organic molecules at the FUV-irradiated edge of the Horsehead nebula. Using ALMA 7m interferometry combined with IRAM 30m single-dish data, we have mapped species such as \CO{}, \isocyanicacid{}, \cyanoacetylene{}, \methylcyanide{}, \ketene{}, and \acetaldehyde{} for the first time in this source, achieving an angular resolution of approximately $\sim$$15^{\prime\prime}$. Our analysis also incorporates deep single-dish observations of \CO{} and C$^{18}$O from the WHISPER survey, along with PdBI maps of \formaldehyde{} and \methanol{} from \citet{Guzman2013}. Integrated line intensity profiles were extracted along the direction of the FUV radiation, and radiative transfer modeling was used to constrain the gas physical conditions and molecular column densities. The main results and conclusions are summarized as follows:

\begin{enumerate}
    \item The column density of \CO{} in the Horsehead nebula has been constrained for the first time. Using previous estimates of $N(\mathrm{H_2})$, an empirical \CO{}-to-H$_2$ conversion factor of $X_{\mathrm{C^{17}O}}= N(\mathrm{H_2})/N(\mathrm{C^{17}O})\sim(5 \pm 1)\times10^{6}$ is obtained. The newly derived ratio $N(\mathrm{C^{18}O})/N(\mathrm{C^{17}O}) \sim 3.5$ further confirms that this $X_{\mathrm{C^{17}O}}$ value is consistent with previous estimates from other CO isotopologues.
    \item Application of the $X_{\mathrm{C^{17}O}}$ factor to the $N(\mathrm{C^{17}O})$ profiles yields empirical $n_{\mathrm{H_2}}$ profiles, corroborating the presence of a steep density gradient at the edge of the Horsehead nebula, in agreement with prior models and observations.
    \item Simultaneous fitting of multiple $K$-ladder transitions of \methylcyanide{} enabled the determination of empirical $T_{\mathrm{kin}}$ profiles, confirming previous temperature estimates at both the dense core and PDR positions obtained from various independent tracers.
    \item From the $n_{\mathrm{H_2}}$ and $T_{\mathrm{kin}}$ profiles, we derived an empirical profile of $P_{\mathrm{th}}$. The high thermal pressure values, together with their relatively small variations across the distance from the FUV radiation source, confirm the compressed and isobaric nature of the Horsehead PDR.
    \item Except for \cyanoacetylene{} and \methanol{}, most of the analyzed molecules—specifically H$_2$CO, CH$_2$CO, CH$_3$CHO, HNCO, and CH$_3$CN—show a clear increase in column density toward the PDR, suggesting that exposure to FUV irradiation favors their gas-phase abundance, contrary to classical PDR expectations. In the Horsehead, this likely points to a significant contribution from grain surface chemistry, given that \cyanoacetylene{} is not efficiently formed on grains, while \methanol{}, although formed on grains, is not efficiently desorbed.
    \item Given that atomic C is expected to be abundant in PDRs due to CO photodissociation, and considering the widespread detection of O- and N-bearing COMs, the Horsehead nebula serves as a compelling example of how surface diffusion of atomic C on grains could enhance chemical complexity in the ISM, as recently suggested by laboratory experiments. Atomic C may contribute either by directly forming COMs on grain surfaces or by producing precursors that are subsequently desorbed, driving further chemical evolution in the gas phase.
\end{enumerate}

Overall, compared to regions exposed to stronger FUV radiation fields, such as the Orion Bar, the Horsehead PDR appears to offer an optimal combination of temperature and UV intensity to promote the efficient formation of COMs through a complex interplay of gas-phase and grain-surface chemistry. Disentangling the relative contributions of these pathways will require higher angular resolution observations and/or \textit{James Webb Space Telescope} (JWST) ice studies toward stars located behind the Horsehead. Our results further suggest that similar chemical processes may operate in other cold, UV-irradiated environments with abundant CO and H$_2$O ices, such as the molecular layers of protoplanetary disks. However, additional chemical modeling and new observations will be essential to assess the broader applicability of these findings to other sources.

\begin{acknowledgements}
The authors thank the referee for the constructive comments that improved the content of this work. This paper makes use of the following ALMA data: ADS/JAO.ALMA\#2016.2.00027.S. ALMA is a partnership of ESO (representing its member states), NSF (USA) and NINS (Japan), together with NRC (Canada), NSTC and ASIAA (Taiwan), and KASI (Republic of Korea), in cooperation with the Republic of Chile. The Joint ALMA Observatory is operated by ESO, AUI/NRAO and NAOJ. The National Radio Astronomy Observatory is a facility of the National Science Foundation operated under cooperative agreement by Associated Universities, Inc. Data reduction was carried out at the IRAM ARC node. C.H.-V. acknowledges support from the National Agency for Research and Development (ANID) -- Scholarship Program through the Doctorado Nacional grant no. 2021-21212409. C.H.-V. and 
V.V.G. acknowledge support from ANID -- Millennium Science Initiative Program -- Center Code NCN2024\_001. V.V.G also gratefully acknowledges support from FONDECYT Regular 1221352, and ANID CATA-BASAL project FB210003. J.P. and M.G. acknowledge support by the French Agence Nationale de la Recherche through the DAOISM grant ANR-21-CE31-0010 and by the Thematic Action “Physique et Chimie du Milieu Interstellaire” (PCMI) of INSU Programme National “Astro”, with contributions from CNRS Physique \& CNRS Chimie, CEA, and CNES. K.T.W. acknowledges support from the European Research Council (ERC) under the European Union's Horizon 2020 research and innovation programme (Grant agreement no. 883867, project EXWINGS). J.R.G. thanks the Spanish MCINN for funding support under grant PID2023-146667NB-I00.
\end{acknowledgements}

\bibliographystyle{aa} 
\bibliography{main}

\begin{appendix} 

\onecolumn

\section{Observational details}

Table~\ref{table:aca} summarizes the observational details of the spectral setup used in the ALMA 7m observations.

\begin{table}[h!]
\caption{\label{table:aca}Spectral setup of the ALMA 7m observations}
\centering
\begin{tabular}{cccccc}
\hline\hline
BB & $\nu_{\rm rest}$ & Species & $\Delta\nu$ & BW    & $\nu_{\rm TDM}$ \\
   & (GHz)            &         & (kHz)       & (MHz) & (GHz)           \\
\hline
1 & 98.00 & $l$-C$_3$H & 30.5 & 62.5 & 98.32 \\
  & 98.86 & CH$_3$CHO  & 30.5 & 62.5 & \\ 
\hline
2 & 100.09 & CH$_2$CO, HC$_3$N & 30.5 & 62.5 & 100.28 \\
  & 100.53 & CH$_3$NC  & 30.5 & 62.5 & \\ 
\hline
3 & 109.91 & HNCO      & 30.5 & 62.5 & 110.32 \\
  & 110.38 & CH$_3$CN  & 30.5 & 62.5 & \\ 
\hline
4 & 112.25 & CH$_3$CHO & 244 & 250 & 112.32 \\
  & 112.36 & C$^{17}$O & 244 & 250 & \\
  & 112.70 & Continuum & 244 & 500 & \\
\hline
\end{tabular}
\tablefoot{Columns from left to right: Baseband number, rest frequency (GHz) at the center of the spectral window, targeted species, channel spacing (kHz), bandwidth (MHz), and the central frequency (GHz) of the TDM spectral window used in calibration.}
\end{table}

\clearpage

\section{Bayesian fitting results}\label{App:Best-fit}

\subsection{WHISPER CO isotopologues observations}\label{subApp:WHISPER}

\begin{figure*}[b!]
\centering
\includegraphics[width=0.9\linewidth]{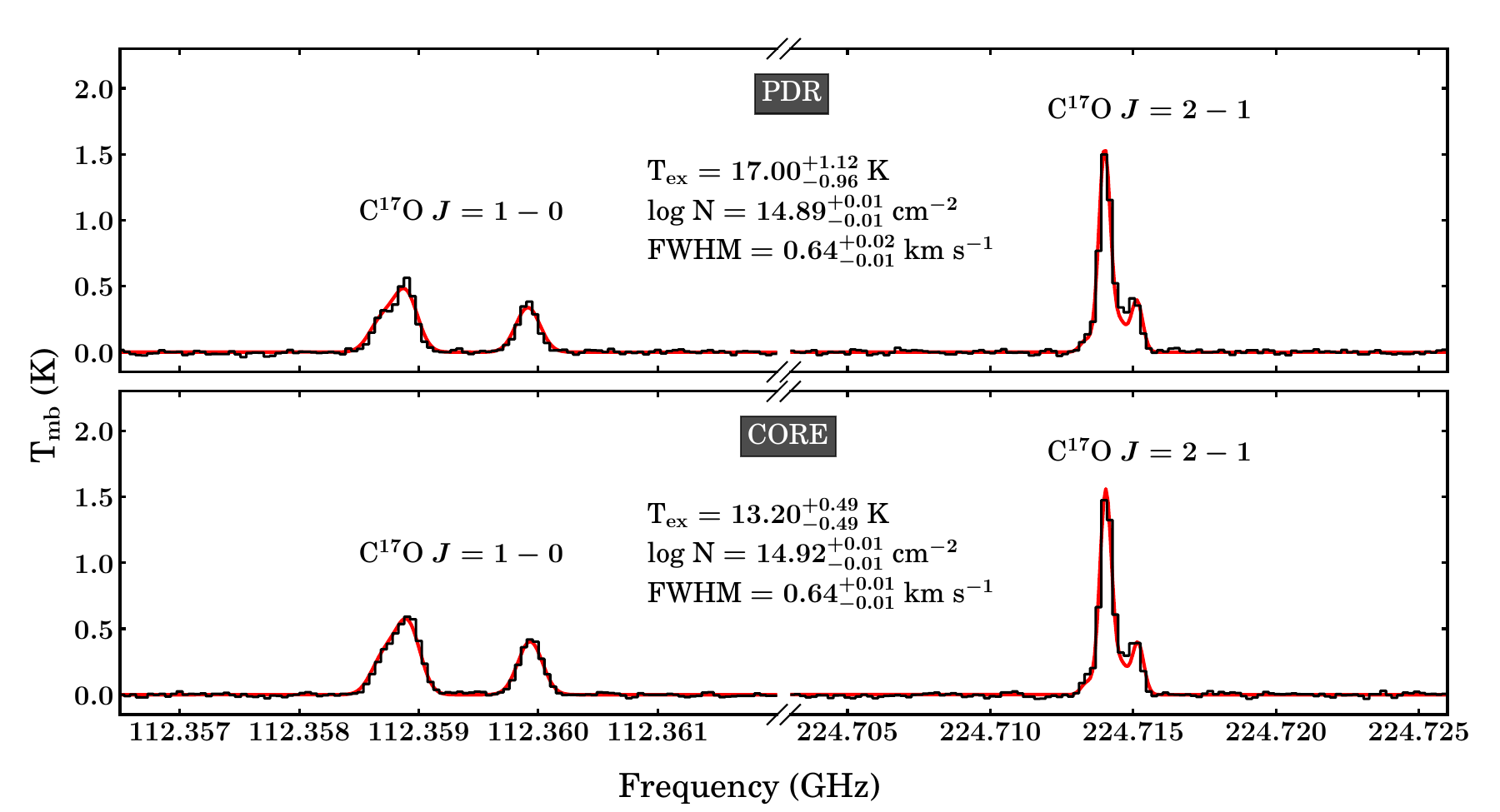}
\includegraphics[width=0.9\linewidth]{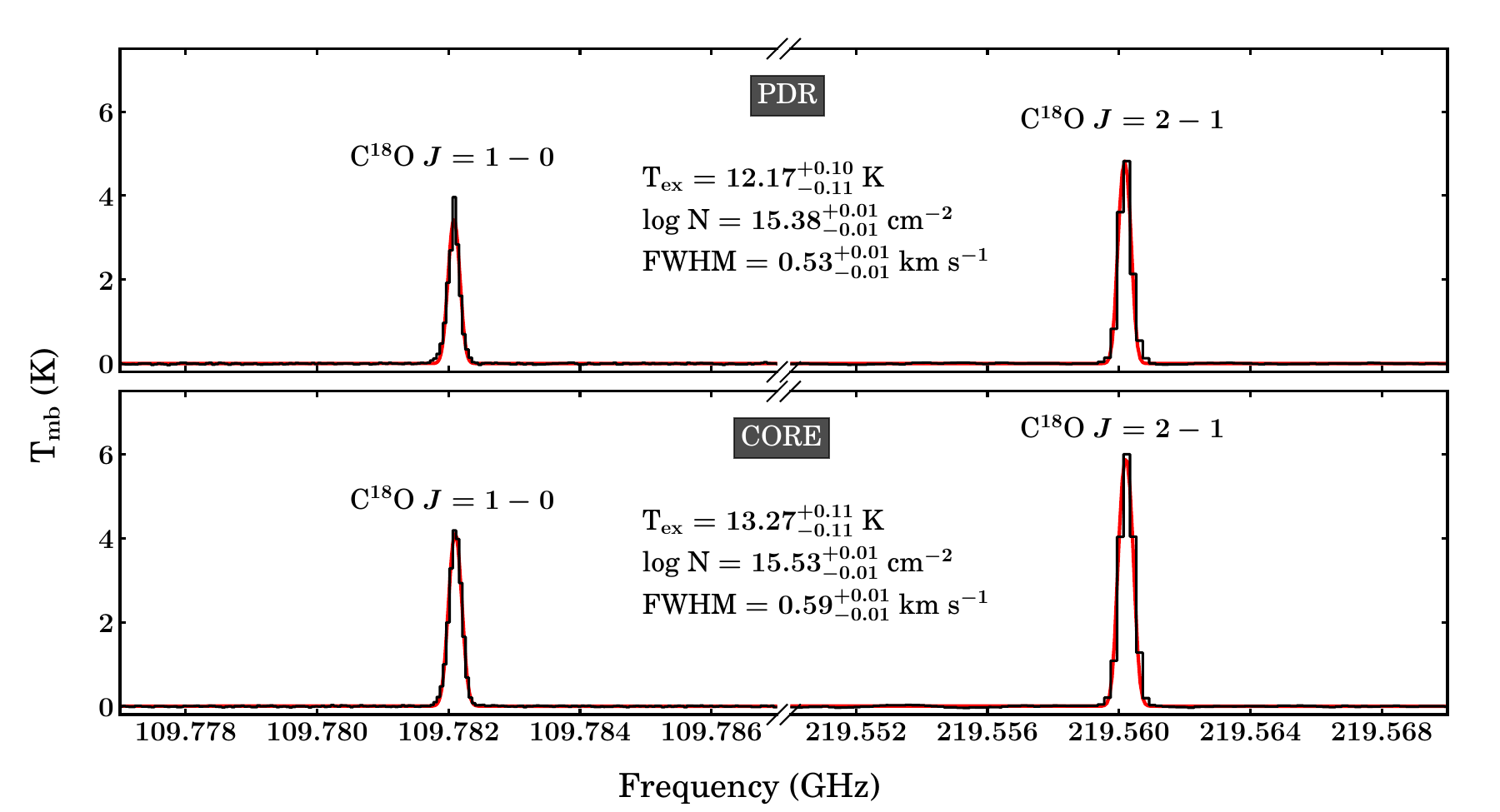}
\caption{C$^{17}$O (top panels) and C$^{18}$O (bottom panels) spectra at the PDR position ($\delta x \approx 15\farcs1, \:\delta y \approx 1\farcs2$) and dense core position ($\delta x \approx 44\farcs7, \:\delta y \approx 16\farcs5$), as observed in the Horsehead WHISPER survey. The best-fit models are overlaid in red. The best-fit excitation conditions and associated uncertainties, shown in the middle of each panel, correspond to the 16th, 50th, and 84th percentiles of the posterior distributions.}
\label{fig:hh-WHISPER-fits}
\end{figure*}

Figure~\ref{fig:hh-WHISPER-fits} presents the best-fit models accounting for the hyperfine structure of the \CO{} (top panels) and C$^{18}$O (bottom panels) $J=1-0$ and $2-1$ rotational transitions, based on the WHISPER data analysis described in Sect.~\ref{Subsubsec:conversion-factor}. The best-fit parameters and their statistical uncertainties for the PDR and dense core positions are shown in the middle of each panel. Averaging the results for the PDR and dense core yields $T_{\mathrm{ex}}(\mathrm{C^{17}O}) \sim 15$~K and $N(\mathrm{C^{18}O})/N(\mathrm{C^{17}O}) \sim 3.6$, which are the values adopted in Sect.~\ref{Subsubsec:conversion-factor} for the posterior analysis of the ALMA 7m+30m combined dataset.

\subsection{ALMA 7m+30m \texorpdfstring{C$^{17}$O}{C17O} observations}\label{subApp:C17O}

\begin{figure*}[h!]
\centering
\includegraphics[width=0.9\linewidth]{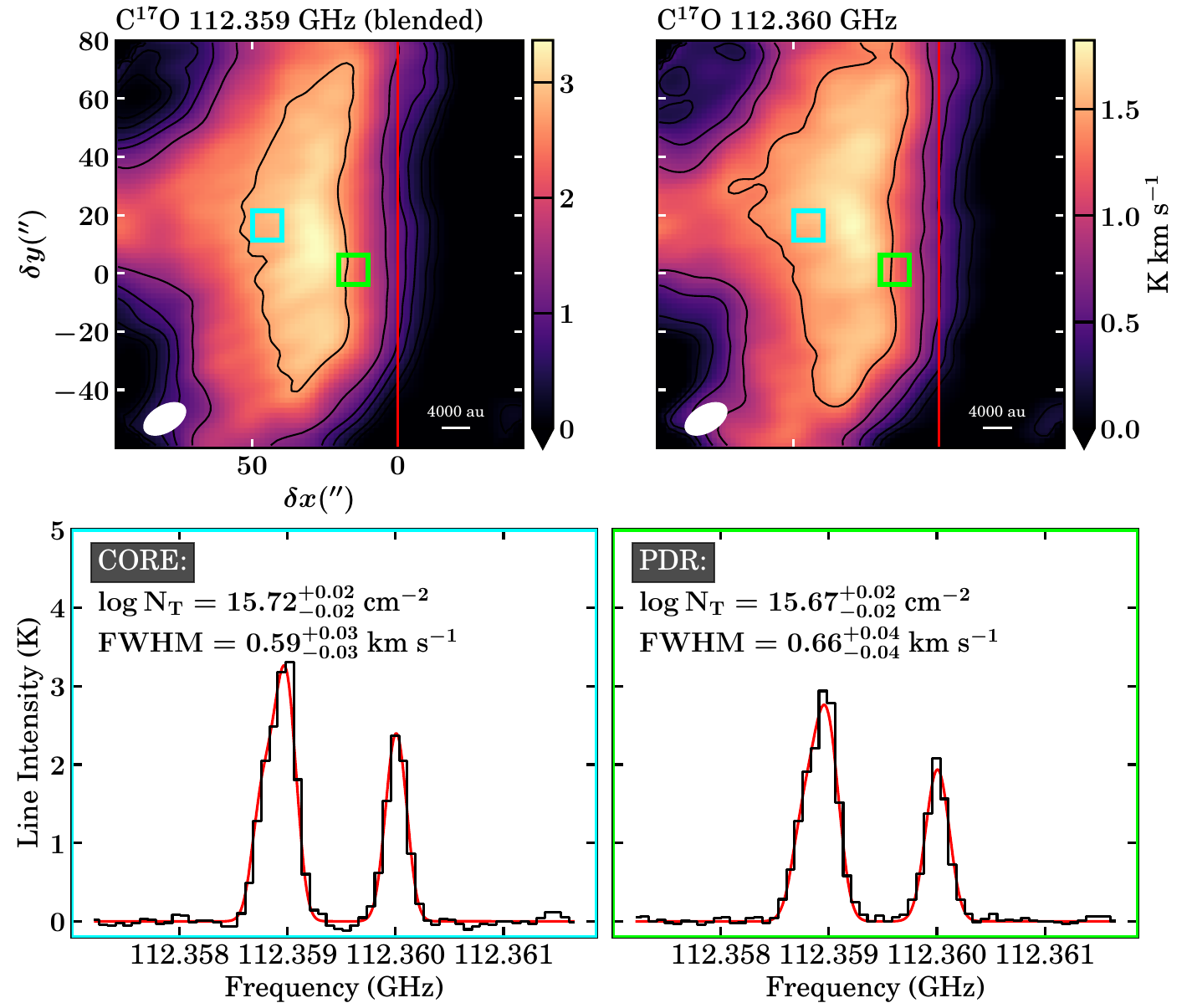}
\caption{\textit{Top row:} ALMA 7m+30m zeroth-moment maps of the hyperfine components of the \CO{} $J=1-0$ rotational transition resolved at the Horsehead edge. The rotation angle and red vertical line are the same as in Fig.~\ref{fig:hh-coms-grid}. The contours are $[5, 10, 20, 30, 50, 100]\times\sigma$, where $\sigma$ is the zeroth-moment rms listed in Table~\ref{table:obs-params}. The dense core and PDR positions are marked with cyan and green squares, respectively, indicating the regions where the spectra were extracted. The beam size and a scale bar indicating $4000$~au are shown in the bottom left and bottom right corner, respectively, of each panel. \textit{Bottom row:} \CO{} spectra extracted at the dense core and PDR positions, shown using matching colors to the map markers. The best-fit models are overlaid in red. The best-fit excitation conditions and associated uncertainties, presented in the top left corner of each panel, correspond to the 16th, 50th, and 84th percentiles of the posterior distributions.}
\label{fig:hh-c17o-grid}
\end{figure*}

Figure~\ref{fig:hh-c17o-grid} presents zeroth-moment maps of the hyperfine components of the \CO{} $J=1-0$ rotational transition from the combined ALMA 7m+30m dataset. Only two maps are displayed, as the two hyperfine components at $112.359$~GHz are blended (see Table~\ref{table:app-moldata}). The green and cyan squares mark the PDR and dense core positions, respectively, where the spectra were extracted for the calibration of the \CO{}-to-H$_2$ conversion factor.

Below the maps, the corresponding spectra and best-fit models are displayed, with the derived parameters and their associated uncertainties indicated, following a similar format as in Fig.~\ref{fig:hh-WHISPER-fits}. By assuming $T_{\mathrm{ex}}(\mathrm{C^{17}O}) = 15$ K, we derive column densities of $N(\mathrm{C^{17}O}) = (4.7 \pm 0.2)\times10^{15}$ cm$^{-2}$ for the PDR and $N(\mathrm{C^{17}O}) = (5.2 \pm 0.2)\times10^{15}$ cm$^{-2}$ for the dense core. These values are slightly higher than those obtained with WHISPER, likely due to beam dilution. As a sensitivity test, we repeated the fitting procedure using the $T_{\mathrm{ex}}$ values estimated from WHISPER at both positions (see Fig.~\ref{fig:hh-WHISPER-fits}), finding no substantial changes in the derived $N(\mathrm{C^{17}O})$ or FWHM values.

These results, together with $N(\mathrm{H_2})$ estimates from \citet{Guzman2014}, were used to calibrate the \CO{}-to-H$_2$ conversion factor discussed in Sect.~\ref{Subsubsec:conversion-factor}. Equivalent plots were generated (but are not shown) for the remaining $\delta x$ positions at the heights of the PDR ($\delta y \sim 1\farcs2$) and dense core ($\delta y \sim 16\farcs5$), enabling the computation of the \nH{} profiles as described in Sect.~\ref{Subsubsec:gas-density}.

\clearpage

\subsection{ALMA 7m+30m \texorpdfstring{CH$_3$CN}{CH3CN} observations}\label{subApp:CH3CN}

\begin{figure*}[h!]
\centering
\includegraphics[width=1.0\linewidth]{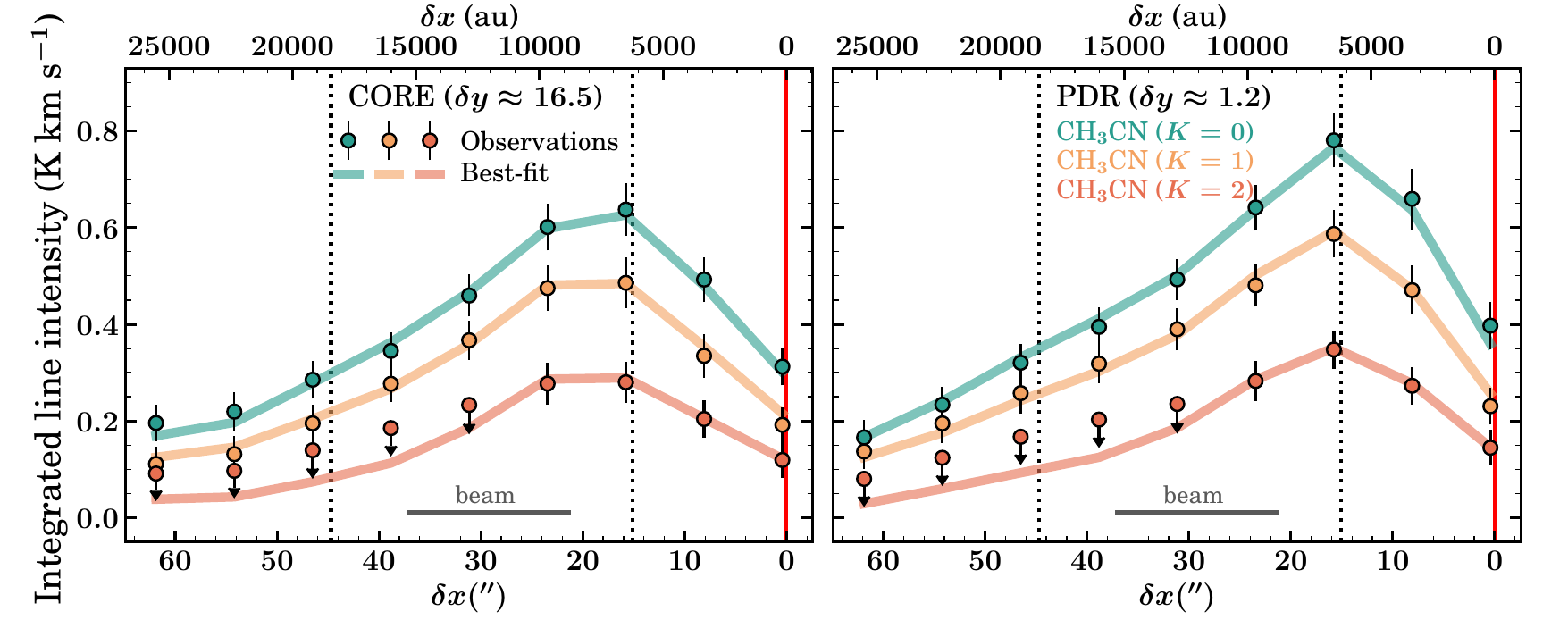}
\caption{Integrated intensity profiles of the \methylcyanide{} $J=6-5$ ($K=0,1,$ and $2$) transitions at the $\delta y$ positions of the dense core (left panel) and the PDR (right panel), as obtained from the ALMA 7m+30m observations. Each color corresponds to a different $K$-ladder transition, as indicated by the labels in the right panel. The observed data points are shown as circles, and the best-fit models are overlaid as solid lines in the corresponding colors. The red vertical line, the dotted vertical lines, and the horizontal gray bar are the same as in the left panels of Fig.~\ref{fig:hh-coms-profiles}.}
\label{fig:hh-ch3cn-fit}
\end{figure*}

Figure~\ref{fig:hh-ch3cn-fit} presents the best-fit models to the integrated intensity profiles of the various \methylcyanide{} $J=6-5$ ($K=0,1,$ and $2$) transitions, extracted at the vertical positions of the dense core (left panel) and the PDR (right panel), following the procedure described in Sect.~\ref{Subsubsec:gas-temperature}. Near the PDR ($\delta x < 30^{\prime\prime}$), the $K=0,1,$ and $2$ transitions were jointly fit. However, toward the dense core ($\delta x \geq 30^{\prime\prime}$), the integrated intensity of the $K=2$ transition was used only as an upper limit due to observed line profile asymmetries and broader line widths relative to the other $K$-ladder transitions. As the $K=2$ line corresponds to the highest $E_{u}$ among the analyzed \methylcyanide{} transitions (see Table~\ref{table:app-moldata}), we speculate that its emission does not originate in the cold dense core, but rather arises from warmer gas along the line of sight, a phenomenon previously reported in the Horsehead for tracers such as CS and HCO \citep{Goicoechea2006,Gerin2009}. It is also plausible that this warmer, more diffuse component has different kinematics compared to the quiescent dense core, which could explain not only the line broadening but also the asymmetric deviations from a purely Gaussian line profile.

\clearpage

\section{Molecular data} \label{App:Mol-data}

Table~\ref{table:app-moldata} summarizes the spectroscopic constants and the critical densities for all the molecular lines analyzed throughout this work.

\begin{table*}[h!]
\caption{Molecular data of the transitions analyzed in this work.}
\label{table:app-moldata}
\centering
\begin{tabular}{l c c c c c c}
\hline\hline
Species & Transition & Frequency & $E_{u}$ & $\log_{10}A_{ul}$ & $g_u$ & $n_{\mathrm{crit}}$\tablefootmark{a}\\
 &  & (GHz) & (K) & (s$^{-1}$) &  & (cm$^{-3}$) \\
\hline
C$^{17}$O & $J=1-0$, $F=3/2-5/2$ & 112.359 & 5.39 & -7.174 & 4 & $(2.0-2.1)\times10^{3}$\\
 & $J=1-0$, $F=7/2-5/2$ & 112.359 & 5.39 & -7.174 & 8 & $(2.0-2.1)\times10^{3}$ \\
 & $J=1-0$, $F=5/2-5/2$ & 112.360 & 5.39 & -7.174 & 6 & $(2.0-2.1)\times10^{3}$ \\
 & $J=2-1$, $F=3/2-5/2$ & 224.713 & 16.18 & -6.715 & 4 & $(1.0-1.1)\times10^{4}$\\
 & $J=2-1$, $F=5/2-5/2$ & 224.714 & 16.18 & -6.407 & 6 & $(1.0-1.1)\times10^{4}$\\
 & $J=2-1$, $F=7/2-5/2$ & 224.714 & 16.18 & -6.384 & 8 & $(1.0-1.1)\times10^{4}$\\
 & $J=2-1$, $F=9/2-7/2$ & 224.714 & 16.18 & -6.192 & 10 & $(1.0-1.1)\times10^{4}$\\
 & $J=2-1$, $F=1/2-3/2$ & 224.714 & 16.18 & -6.192 & 2 & $(1.0-1.1)\times10^{4}$\\
 & $J=2-1$, $F=3/2-3/2$ & 224.715 & 16.18 & -6.347 & 4 & $(1.0-1.1)\times10^{4}$\\
 & $J=2-1$, $F=5/2-7/2$ & 224.715 & 16.18 & -7.292 & 6 & $(1.0-1.1)\times10^{4}$\\
 & $J=2-1$, $F=7/2-7/2$ & 224.715 & 16.18 & -6.639 & 8 & $(1.0-1.1)\times10^{4}$\\
 & $J=2-1$, $F=5/2-3/2$ & 224.715 & 16.18 & -6.699 & 6 & $(1.0-1.1)\times10^{4}$\\
C$^{18}$O & $J=1-0$ & 109.782 & 5.27 & -7.203 & 3 & $(1.8-1.9)\times10^{3}$\\
 & $J=2-1$ & 219.560 & 15.81 & -6.221 & 5 & $(0.9-1.0)\times10^{4}$\\
H$_2$CO & $J_{K_a,K_c}=2_{02}-1_{01}$ & 145.603 & 10.48 & -4.107 & 5 & $(0.9-1.1)\times10^{6}$ \\
CH$_3$OH & $J_{K} = 3_{-1}-2_{-1}$ (E) & 145.097 & 19.51 & -4.960 & 7 & $(8.4-9.1)\times10^{4}$ \\
CH$_2$CO\tablefootmark{b} & $J_{K_a,K_c} = 5_{15}-4_{14}$ & 100.095 & 27.46 & -4.988 & 33 & $-$\\
CH$_3$CHO\tablefootmark{b} & $J_{K_a,K_c} = 5_{14}-4_{13}$ (E) & 98.863 & 16.59 & -4.508 & 22 & $-$\\
 & $J_{K_a,K_c} = 6_{16}-5_{15}$ (E) & 112.255 & 21.21 & -4.330 & 26 & $-$\\
 & $J_{K_a,K_c} = 6_{16}-5_{15}$ (A) & 112.249 & 21.13 & -4.331 & 26 & $-$\\
HNCO & $J_{K_a,K_c} = 5_{05}-4_{04}$ & 109.906 & 15.82 & -4.756 & 11 & $(2.4-2.5)\times10^{5}$ \\
HC$_3$N & $J=11-10$ & 100.076 & 28.82 & -4.110 & 23 & $(1.5-1.7)\times10^{6}$ \\
CH$_3$CN & $J_K = 6_{0}-5_{0}$ (A) & 110.384 & 18.54 & -3.954 & 26 & $(1.9-2.4)\times10^{6}$\\
 & $J_K = 6_{1}-5_{1}$ (E) & 110.381 & 25.69 & -3.966 & 26 & $(2.0-2.5)\times10^{6}$\\
 & $J_K = 6_{2}-5_{2}$ (E) & 110.375 & 47.12 & -4.005 & 26 & $(1.6-2.0)\times10^{6}$\\
\hline
\end{tabular}
\tablefoot{This covers all the lines analyzed throughout this work, including the legacy data described in Sect.~\ref{Subsec:auxlliary-data}. All molecular data are taken from the Cologne Database for Molecular Spectroscopy \citep[CDMS;][]{Endres2016} except for \acetaldehyde{}, which is from the Jet Propulsion Laboratory database \citep[JPL;][]{Pickett1998}. \\
\tablefoottext{a}{For those species with available collisional rates in the Leiden Atomic and Molecular Database \citep[LAMDA;][]{Schoier2005}, the critical density ($n_{\mathrm{crit}}$) is estimated as the density at which the rate of spontaneous radiative de-excitation ($A_{ul}$) equals the rate of collisional de-excitation ($\gamma_{ul}$) for temperatures between $20-60$~K.}\\
\tablefoottext{b}{Although \ketene{} and \acetaldehyde{} lack available collisional rates, their emission is likely sub-thermal ($n_{\mathrm{H}_2} < n_{\mathrm{crit}}$), as suggested by the low \Tex{} values reported by \citet{Guzman2014}, which are adopted in this work.}
}
\end{table*}

\clearpage

\section{Abundances at the PDR and dense core positions}\label{App:N-core}

\begin{figure*}[h!]
\centering
\includegraphics[width=1.0\linewidth]{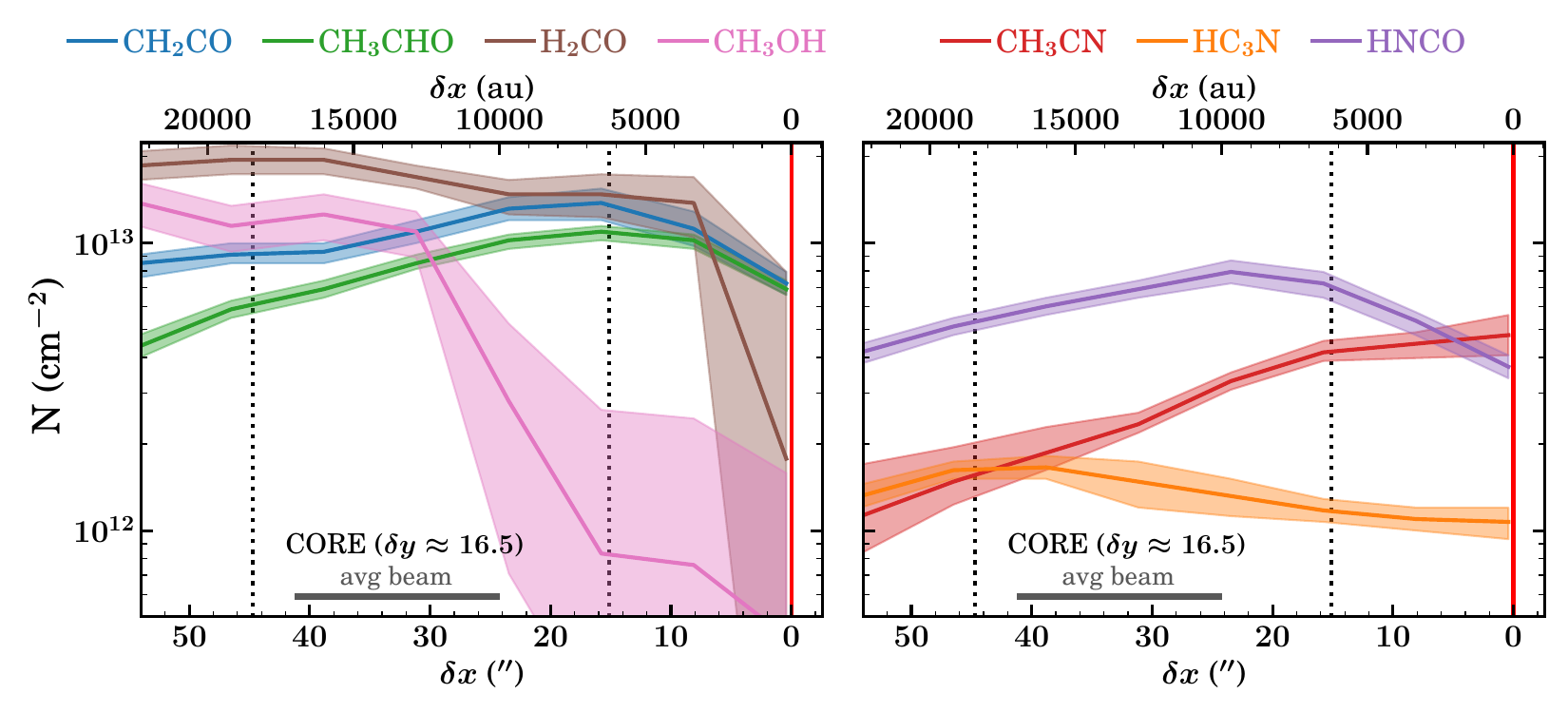}
\caption{Same as Fig.~\ref{fig:hh-logN-profiles}, but at the $\delta y$ position of the dense core.}
\label{fig:hh-logN-profiles-CORE}
\end{figure*}

Figure~\ref{fig:hh-logN-profiles-CORE} presents the column density profiles derived at the vertical position of the dense core, analogous to those shown for the vertical position of the PDR in Sect.~\ref{Subsubsec:column-densities}. For comparison with the Orion Bar PDR, we extracted the column density values specifically at the $\delta x$ positions of the dense core ($\delta x \approx 44\farcs7$, from Fig.~\ref{fig:hh-logN-profiles-CORE}) and the PDR ($\delta x \approx 15\farcs1$, from Fig.~\ref{fig:hh-logN-profiles}) to estimate their abundances with respect to total hydrogen nuclei. The resulting values, together with those for the Orion Bar reported in the literature, are listed in Table~\ref{table:abundances}.

\begin{table*}[h!]
\def\arraystretch{1.5}
\caption{Abundances with respect to total hydrogen nuclei ($N(\mathrm{X})/N_{\mathrm{H}}$, in units of $10^{-10}$).}
\label{table:abundances}
\centering
\begin{tabular}{c c c | c}          
\hline\hline
Species & \multicolumn{2}{c}{Horsehead\tablefootmark{a}} & \multicolumn{1}{|c}{Orion Bar\tablefootmark{b}} \\
& PDR & Dense core & PDR \\   
\hline
\formaldehyde{} & $3.4^{+0.5}_{-0.5}$ & $3.0^{+0.4}_{-0.3}$ & $9.0$ \\
\methanol{} & $0.4^{+0.7}_{-0.3}$ & $1.8^{+0.3}_{-0.3}$ & $5.0$ \\
\ketene{} & $4.7^{+0.3}_{-0.3}$ & $1.4^{+0.1}_{-0.1}$ & $0.9$  \\
\acetaldehyde{} & $3.9^{+0.2}_{-0.1}$ & $0.9^{+0.1}_{-0.1}$ & $0.8$ \\
\methylcyanide{} & $1.3^{+0.1}_{-0.1}$ & $0.2^{+0.1}_{-0.1}$ & $0.2$  \\
\cyanoacetylene{} & $0.3^{+0.1}_{-0.1}$ & $0.3^{+0.1}_{-0.1}$ & $0.1$  \\
\isocyanicacid{} & $2.5^{+0.2}_{-0.1}$ & $0.8^{+0.1}_{-0.1}$ & $0.2$ \\
\hline
\end{tabular}
\tablefoot{The total hydrogen nuclei column density is estimated as $N_{\mathrm{H}} = N(\mathrm{X})/(N(\mathrm{H})+2N(\mathrm{H}_2))$.\\ 
\tablefoottext{a}{This work. Abundances calculated using $N_{\mathrm{H}} = 3.8\times10^{22}$~cm$^{-2}$ (PDR) and $N_{\mathrm{H}} = 6.4\times10^{22}$~cm$^{-2}$ (dense core).} \\
\tablefoottext{b}{Values from \cite{Cuadrado2017}. Abundances calculated using $N_{\mathrm{H}} = 6.3\times10^{22}$~cm$^{-2}$.}
}
\end{table*}

\end{appendix}
\end{document}